\RequirePackage{tikz}

\documentclass[12pt]{article}

\usepackage{graphicx}%
\usepackage{multirow}%
\usepackage{amsmath,amssymb,amsfonts}%
\usepackage{amsthm}%
\usepackage{mathrsfs}%
\usepackage[title]{appendix}%
\usepackage{xcolor}%
\usepackage{relsize} 

\theoremstyle{thmstyleone}%

\theoremstyle{thmstyletwo}%
\newtheorem{remark}{Remark}%

\theoremstyle{thmstylethree}%

\usepackage{upgreek,bm}%

\newcommand\DT[1]{\frac{{\rm D}#1}{{\rm D}t}}
\newcommand\Dt[1]{\mathchoice
                 {{\buildrel{\hspace*{.1em}\text{\LARGE.}}\over{#1}}}
                 {{\buildrel{\hspace*{.1em}\text{\LARGE.}}\over{#1}}}
                 {{\buildrel{\hspace*{.1em}\text{\Large.}}\over{#1}}}
                 {{\buildrel{\hspace*{.1em}\text{\large.}}\over{#1}}}}
\newcommand\DDt[1]{\mathchoice
   {{\buildrel{\hspace*{.1em}\text{\Large.\hspace*{-.1em}.}}\over{#1}}}
   {{\buildrel{\hspace*{.1em}\text{\large.\hspace*{-.1em}.}}\over{#1}}}
   {{\buildrel{\hspace*{.1em}\text{\large.\hspace*{-.1em}.}}\over{#1}}}
   {{\buildrel{\hspace*{.1em}\text{\large.\hspace*{-.1em}.}}\over{#1}}}}
\newcommand\DDDt[1]{\mathchoice
   {{\buildrel{\hspace*{.1em}\text{\Large.\hspace*{-.1em}\Large.\hspace*{-.1em}.}}\over{#1}}}
   {{\buildrel{\hspace*{.1em}\text{\large.\hspace*{-.1em}\large.\hspace*{-.1em}.}}\over{#1}}}
   {{\buildrel{\hspace*{.1em}\text{\large.\hspace*{-.1em}\large.\hspace*{-.1em}.}}\over{#1}}}
   {{\buildrel{\hspace*{.1em}\text{\large.\hspace*{-.1em}\large.\hspace*{-.1em}.}}\over{#1}}}}
\newcommand\OLD[1]{\mathchoice
               {{\buildrel{\hspace*{.1em}{_{\,\boldsymbol\triangledown}}}\over{#1}}}
               {{\buildrel{\hspace*{.1em}{_{\,\boldsymbol\triangledown}}}\over{#1}}}
               {{\buildrel{\hspace*{.1em}{\boldsymbol\triangledown}}\over{#1}}}
               {{\buildrel{\hspace*{.1em}{\boldsymbol\triangledown}}\over{#1}}}}
\newcommand\ZJ[1]{\mathchoice
                 {{\buildrel{\hspace*{.1em}{_{\,\boldsymbol\circ}}}\over{#1}}}
                 {{\buildrel{\hspace*{.1em}{_{\,\boldsymbol\circ}}}\over{#1}}}
                 {{\buildrel{\hspace*{.1em}{\boldsymbol\circ}}\over{#1}}}
                 {{\buildrel{\hspace*{.1em}{\boldsymbol\circ}}\over{#1}}}}
\newcommand{\eq}[1]{(\ref{#1})}
\def\d{{\rm d}}
\newcommand{\pl}{\partial}
\def\Kv{D}
\def\vv{\bm v}
\def\XX{\bm X}
\def\xx{\bm x}
\def\vv{\bm v}

\def\ee{\bm e}
\def\FF{{\bm F}}
\def\LL{{\bm L}}

\def\TT{\bm T}
\def\BB{{\bm B}}
\def\WW{{\bm W}}
\def\bbD{\mathbb D}
\def\bbC{\mathbb C}
\def\bbI{\mathbb I}
\def\EE{\bm E}

\def\EL{_{_\text{\sc el}}}
\def\IN{_{_\text{\sc in}}}
\newcommand{\lineunder}[2]{\LU{\begin{array}[t]{c}\underbrace{#1}\vspace*{.5em}\end{array}}{\mbox{\footnotesize\rm #2}}}
\newcommand{\LU}[2]{\begin{array}[t]{c}#1\vspace*{-1em}\\_{#2}\end{array}}

\usepackage{psfrag}    
\usepackage{tikz}  
\usetikzlibrary{decorations.pathmorphing}
\usetikzlibrary{patterns}
\usepackage{xcolor} 
\usepackage{pgfplots}
\usetikzlibrary{calc}
\newcommand{\R}{{\mathbb R}}

\begin{document}

\noindent
{\Large\bf
  Some gradient theories in linear visco-elastodynamics\\[.3em]towards
  dispersion and attenuation of waves\\[.3em]in relation
  to large-strain models}

\bigskip\bigskip\bigskip

\noindent
{\large Tom\'a\v s Roub\'\i\v cek}

\bigskip\bigskip\bigskip

{\small
\noindent
{\bf Abstract}.
Various spatial-gradient extensions
  of standard viscoelastic rheologies of the Kelvin-Voigt,
  Maxwell's, and Jeffreys' types are analyzed in linear
  one-dimensional situations as far as the propagation of
  waves and their dispersion and attenuation. These gradient
  extensions are then presented in the large-strain nonlinear variants 
  where they are sometimes used rather for purely analytical
  reasons either in the Lagrangian or the Eulerian formulations
  without realizing this wave-propagation context. The interconnection
  between these two modeling aspects is thus revealed in particular
  selected cases.

\medskip


  \medskip


}

\bigskip

\baselineskip=16pt

\section{Introduction}\label{sec1}

Propagation of {\it elastic waves} in solid or in (visco)elastic fluidic media
is an important part of continuum mechanics with important applications
in mechanical engineering, material science, seismology, etc. This has been vastly
investigated both in linear and in nonlinear regimes, and both in simple elastic
continua and in various ``complex'' continua, supported by various
microscopical ideas, specifically microstructured media
\cite{BeEnBe11WMSU,BeYiSc20WDMS,EngBer15RMMD,EBPB05WMMD,Mind64MSLE},
micropolar or micromorphic media \cite{Erin99MFT,MNAB17RMIE}, heterogeneous media
\cite{FiChNa02NDMW}, viscoelastic media (referred below), porous media
\cite{MuGuLe10SWAD}, granular materials \cite{Wilm99WPGM}, active media, etc. It
has resulted in  many visco-elastic rheological models of both the solid and the
fluidic types used in continuum-mechanical modelling. Besides purely elastic
 Hooke-type rheology, the
basic options are Kelvin-Voigt (solid) and Maxwell (fluidic). The simplest combination
then leads to the standard solid (Zener or Poynting-Thomson) or Jeffreys' fluid (also
called anti-Zener). Of course, further more intricate combinations are also popular
but out of the scope of this paper.

These standard rheologies involving the strains and the strain rates are often enhanced
by some higher spatial gradients or some additional memory-like time effects.
We will focus on the former option. Many higher-gradient visco-elastic models can be found
in literature. In particular, they can serve well for fitting the dispersion and attenuation
(absorption) of elastic waves to particular experimental observation or for facilitation
rigorous proofs of a well-posedeness of large-strain variants of such models. These
very different aspects have been scrutinized respectively in different communities
of mechanical engineers and physicists (like seismologists or material scientists)
or applied mathematicians (analysts or numericians). As a result, the relation
between these two aspects is not addressed in literature.

This paper aims to reveal some connections between these two aspects
of selected gradient theories in basic visco-elastic models. The velocity
dispersion and attenuation of waves is quite impossible to analyze effectively
in nonlinear situations and, naturally, we will analyze it in linear models
in one-dimensional situations. After presenting it for three standard rheologies
of simple linear materials in Section~\ref{sec2}, we will analyze selected
nonsimple linear materials in Section~\ref{sec3}; the adjective ``nonsimple''
refers to various higher-order gradients involved in models of such materials.
Some surprising effects are presented in the perspective usage of these gradient
theories in nonlinear large-strain variants, which is
later briefly surveyed in Section~\ref{sec4}.  This represents, together with
the interconnection and certain reflection of the dispersion/attenuation 1-dimensional
linear analysis and the large-strain nonlinear variants, the main asset of this
article. 

Basic types of {\it dispersion} is {\it normal} (the high-frequency waves propagate
 slower  than low-frequency ones)
versus {\it anomalous dispersion} (the high-frequency waves propagate 
faster than low-frequency ones); of course, in terms of the wavelength, the
dependence is the opposite. Besides, some models are nondispersive or with
a general dispersion nonmonotonically dependent on the frequency (or the wavelength).

The other important attribute of wave propagation is their possible {\it attenuation}.
It is roughly quantified by a so-called {\it quality factor}, briefly {\it Q-factor}.
Among various definitions (originally devised rather for oscillating electric
circuits), a mechanically suitable definition is $2\uppi\,\times\,$ratio
between the kinetic or strain stored energy and energy dissipated per cycle.
Like dispersion, also  the  Q-factor may depend on the frequency
(or wavelength) of waves.
A very low Q-factor means that waves cannot propagate at all. Conversely,
Q-factor $+\infty$ means that there is no attenuation and the model is fully
conservative as far as the energy of waves is concerned.
These attributes can be arbitrarily combined  with  various rheological models.

The basic rheological viscoelastic models are often
enhanced by various higher spatial gradients. These spatial-gradient
enhancements are useful also for mathematical
reasons especially in nonlinear situations arising at large
strains, but they can also serve in modelling various internal
length scales and various velocity dispersion and attenuation of elastic waves.
There are many possibilities. Let us
sort some options scrutinized in this paper qualitatively
(together with some standard ``simple'' rheologies) in Table~1.

For completeness, let us mention that another way to build viscoelastic models
is by using various integral operators either in space or in time, cf.\
\cite{BKCK20FVMP,CapMai71LMDA,JelZor23FAZZ,Jira04NTCM,Main10FCWL,TreCox14MPLA}.
We will intentionally focus on rheologies governed by classical
differential equations with higher gradients exclusively in space,
with the aim to enlighten applications of gradient theories used
in large-strain continuum-mechanical models.

\begin{center}
\begin{table}
\caption{Basic classification of the dispersive models in this paper.}
\begin{tabular}{ |c||l|l| } 
 \hline
$_\text{\footnotesize dispersion}\!\!$ $\Large\backslash$ $\!\!^{\text{\footnotesize Q-factor}^{^{^{^{}}}}}$
&\hspace*{.2em}$+\infty$ (conservative)\  &\ \hspace*{6em}$<+\infty\ $\\ 
\hline \hline
normal &  & Kelvin-Voigt \eq{dispersion-} \\ 
 &  & \ \ \ \ \ \ with gradients \eq{dispersion} or \eq{dispersion+}\\
\hline
anomalous & \,\eq{dispersion+} with $\Kv=0$ & Maxwell \eq{telegraph-eq} \\
& & \ \ \ \ \ \ \ with dissipative gradient \eq{hyper-Maxwell-1D} or\!\!\\
& & \ \ \ \ \ \ \  with conservative gradient \eq{hyper-Maxwell-1D-conserve}\\
\hline
general  &   & Jeffreys \eq{Jeffreys-dispers} \\
& & Kelvin-Voigt\\
& & \ \ \ \ \ \ \ with mixed-gradient \eq{dispersion+-}\\
& & \ \ \ \ \ \ \ with stress gradient \eq{Brenner}  and  \eq{Brenner-disip++}\\
& & \ \ \ \ \ \ \ with kinematic gradient \eq{Brenner-disip++}\\
& &  Maxwell
with stress diffusion \eq{Max-stress-diffuse}\\
\hline
nondispersive &\,wave equation \eq{wave-eq}\ & special dissipative gra- \\
& & \hspace*{4.5em}dient Kelvin-Voigt \eq{dispersion+} \\
& & special kinematic gradient \eq{Brenner-disip}\\
 \hline
\end{tabular}
  \end{table}
\end{center}

\section{Linear 1-D simple rheologies}\label{sec2}

We will consider and analyze the 1-dimensional models, which gives also
information for isotropic multidimensional cases where a split into
the volumetric and the deviatoric parts (relevant respectively for the
longitudinal and shear waves) should be made, cf.\ e.g.\ \cite{TreCox14MPLA}.
The departure is the nondispersive, fully conservative elastodynamic model
governed by the wave equation $\varrho\DDt u-{C}u_{xx}^{}=0$
where $u$ is the displacement, $\varrho>0$ is the mass density and ${C}>0$
is the elastic modulus. Equivalently, in the rate formulation
\begin{align}\label{wave-eq}
  \varrho\Dt v={C}e_x\ \ \ \text{ and }\ \ \ \Dt e=v_x\,,
\end{align}
where $e$ is the strain and $v$ a velocity, meaning $e=u_x$ and $v=\Dt u$.
Here the index $x$ means (and will mean) the partial derivative $\pl/\pl x$
 while the dot denotes the partial derivative in time.
In terms of the stress $\sigma={C}e$, it can also be written as the system
\begin{align}\label{wave-eq-alt}
\varrho\Dt v=\sigma_x\ \ \ \text{ and }\ \ \ \Dt\sigma={C}v_x\,,
\end{align}
the linear relation $\sigma={C}e$ being referred as a {\it  Hooke  law}.
The energetics behind this system is based on the stored energy
$\varphi=\varphi(e)=\frac12{C}e^2$ and the kinetic energy
$\frac12\varrho v^2$.

A conventional way to calculate dispersion and attenuation in linear
media is to use the ansatz
\begin{align}\label{dispersion-ansatz-}
 u={\rm e}^{{\rm i}(w t+kx)}
\end{align}
with the angular
frequency $w=\omega+{\rm i}\gamma$ considered complex with 
$\omega,\gamma\in\R$ with physical dimension  rad/s and the real-valued
 angular wave number $k$ with physical dimension  rad/m;
here ${\rm i}=\sqrt{-1}$ denotes the imaginary unit.
Sometimes, an alternative variant to \eq{dispersion-ansatz-}
with a real-valued angular frequency $w$ but complex-valued
 angular  wave number $k$ can be used, cf.\ \cite[Sect.2.3]{Carc15WFRM}.
 Equivalently, in terms of $\lambda=1/k$, 

\begin{align}\label{dispersion-ansatz}
u={\rm e}^{{\rm i}(w t+x/\lambda)}
\end{align}
with the real-valued  angular  wavelength $\lambda$ with physical
dimension meters  per radian (m/rad). One should emphasize that
the actual wavelength (in meters) described by \eq{dispersion-ansatz} is
$2\uppi\lambda$. A certain calculation advantage of working with the
angular wavelength is the correspondence with the angular frequency so that
the speed of the waves described by \eq{dispersion-ansatz} is just
$v=\omega\lambda$. Moreover, writing \eq{dispersion-ansatz} as
$u={\rm e}^{-\gamma t}{\rm e}^{{\rm i}(\omega t+x/\lambda)}$,
reveals that $\gamma$ determines the attenuation of the
 particular monochromatic wave.

Using it for \eq{wave-eq}, i.e.\ using $\varrho\DDt u=-\varrho w^2u$,
and ${C}u_{xx}=-{C}u/\lambda^2$, we can conclude that simply $\gamma=0$ and
$\varrho\omega^2={C}/\lambda^2$, from which we can see the wave velocity
$v=\omega\lambda=\sqrt{{C}/\varrho}$.

\begin{figure}[ht]
\begin{center}
\psfrag{Maxwell}{\small Maxwell}
\psfrag{Kelvin-Voigt}{\small  Kelvin-Voigt}
\psfrag{Jeffreys}{\small  Jeffreys}
\psfrag{C}{\footnotesize ${C}$}
\psfrag{D}{\footnotesize $\Kv$}
\psfrag{D1}{\footnotesize $\Kv_1$}
\psfrag{D2}{\footnotesize $\Kv_2$}
\psfrag{ein}{\footnotesize $e\IN$}
\psfrag{eel}{\footnotesize $e\EL$}
\psfrag{s1}{\footnotesize $\sigma_1$}
\psfrag{s2}{\footnotesize $\sigma_2$}
\psfrag{e}{\footnotesize $e=u_x^{}$}
\hspace*{-.5em}\includegraphics[width=30em]{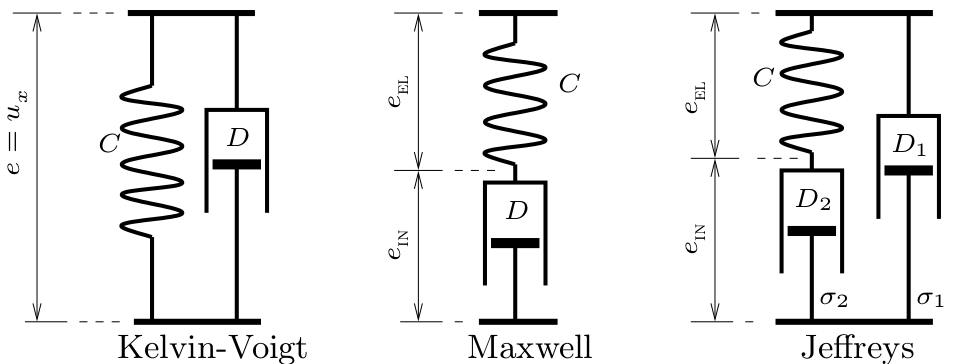}
\end{center}
\vspace*{-.2em}
\caption{
{\sl Schematic 1D-diagrammes of 3 basic rheologies considered in this paper.}
}
\label{fig1}
\end{figure}

We further involve viscosity, using one (or possibly two) linear dashpot element(s)
whose stress/strain response is time-dependent, governed by $\sigma=\Kv\Dt e$ with
the viscosity $\Kv>0$, to be organized in parallel or in series, cf.\ Figure~\ref{fig1}.

\subsection{Kelvin-Voigt visco-elastodynamic model}\label{sec2.1}

Let us start with the basic Kelvin-Voigt rheology. It involves the viscous
dashpot parallel to the elastic element as depicted in Figure~\ref{fig1}-left.
It means an additive decomposition of the stress, which
expands the elastic stress $\sigma={C}e$ in \eq{wave-eq-alt} by a viscous
part as $\sigma=\Kv\Dt e+{C}e$. Thus the
latter equation in \eq{wave-eq-alt} expands as $\Dt\sigma=\Dt v_x+{C}v_x$.
This leads to the dispersive wave equation, i.e.\ in the 1-dimensional variant
\begin{align}\label{dispersion-}
  \varrho \DDt u-\Kv\Dt u_{xx}-{C}u_{xx}^{}=0\,.
\end{align}
Having in mind the ansatz \eq{dispersion-ansatz},
we have $\varrho\DDt u=-\varrho w^2u$,
$\Kv\Dt u_{xx}=-{\rm i}\Kv w u/\lambda^2$,
and ${C}u_{xx}=-{C}u/\lambda^2$, so that the one-dimensional dispersive wave equation
\eq{dispersion-} yields the algebraic condition 
\begin{align}\label{KK-relation-}
\varrho w^2-{\rm i}\,\frac\Kv{\lambda^2}\,w-\frac {C}{\lambda^2}=0\,.
\end{align}
When substituting $w=\omega+{\rm i}\gamma$ so that $w^2=\omega^2-\gamma^2
+2{\rm i}\omega\gamma$, we obtain two algebraic equations for the real and the
imaginary part each, called {\it Kramers-Kronig's relations}
\cite{Kram27DLA,Kron26TDXR}. More specifically, here
\begin{align}\label{KK-relation}
&\varrho(\omega^2-\gamma^2)=\frac {C}{\lambda^2}-
\frac\Kv{\lambda^2}
\gamma\ \ \ \text{ and }\ \ \
2\varrho\gamma=\frac\Kv{\lambda^2}\,.
\end{align}
From the latter equation we can read that $\gamma=\Kv/(2\varrho\lambda^2)$
and then the former equation yields $(\lambda\omega)^2={C}/\varrho
+\gamma^2\lambda^2-\Kv\gamma/\varrho
={C}/\varrho-\gamma^2\lambda^2={C}/\varrho-\Kv^2/(4\varrho^2\lambda^2)$.
Realizing that the speed of
waves is $v=\omega\lambda$, we obtain 
\begin{align}\label{v=v(lambda)}
v=v(\lambda)=\sqrt{\frac {C}\varrho
-\frac{\Kv^2}{4\varrho^2\lambda^2}}\,,
\end{align}
which gives a {\it normal dispersion} for sufficiently long waves, namely having
a length $\lambda>\lambda_\text{\sc crit}$ with the critical  angular 
wavelength $\lambda_\text{\sc crit}=\Kv/(2\sqrt{\varrho {C}})>0$.

Referring to the ansatz  \eq{dispersion-ansatz-} or 
\eq{dispersion-ansatz}, the amplitude decreases ${\rm e}^{-2\uppi\gamma/\omega}$
per one cycle, realizing that one cycle lasts $2\uppi/\omega$ seconds.
 Alternatively, the decay of amplitude is ${\rm e}^{-\gamma/\omega}$
when counted per radian, i.e.\ in units Np/rad with Np denoting so-called ``Neper'',
being an analog of decibels but respecting the Euler number e instead of 10 used
for decibels.  The energy (depending quadratically on the amplitude) thus
decreases by a factor ${\rm e}^{-2\gamma/\omega}$
per one cycle, i.e.\ the loss of energy is $1-{\rm e}^{-2\gamma/\omega}$. 
Thus, not entirely standardly, let us understand the Q-factor as
\begin{align}\label{def-Q-factor}
 \text{Q-factor}\ \sim
\frac1{1-{\rm e}^{-2\gamma/\omega}}\,.
\end{align}
Counting $\omega=v/\lambda$, it also means $1/(1-{\rm e}^{-2\lambda\gamma/v})$.
Thus, taking $\gamma=\Kv/(2\varrho\lambda^2)$ from \eq{KK-relation}, the quality
factor can be understood as
\begin{align}\label{KV-Q-factor}
 \text{Q-factor}\ \sim\ \frac1{1-{\rm e}^{-\Kv/(\varrho\lambda v(\lambda))}}
\ \ \text{ with }\ \ v(\lambda)\ \text{ from \eq{v=v(lambda)}}\,.
\end{align}

\subsection{Maxwellian visco-elastodynamic model}\label{sec2.2}

Let us proceed with the Maxwellian rheology which yields
dispersion of an opposite character, called {\it anomalous}. It uses
the connection of spring 
and dashpot in series, as depicted in Figure~\ref{fig1}-mid,
i.e.\ it employs the Green-Naghdi \cite{GreNag65GTEP}
{\it additive decomposition} of the total strain $e=u_x$ into the elastic
and the inelastic strains, denoted by $e\EL$ and $e\IN$,
respectively. The inelastic strain is in a position
of a so-called {\it internal variable}. The constitutive equations are
\begin{align}\label{Maxewell-internal-par}
 u_x=e\EL+e\IN\ \ \text{ and }\ \sigma={C}e\EL=\Kv\Dt e\IN\,.
\end{align}
From this, we can eliminate $e\IN$ by differentiating
the  Hooke  law as $\Dt\sigma={C}\Dt e\EL$ and the additive decomposition
written in rates using the velocity $v=\Dt u$ as $v_x=\Dt e\EL+\Dt e\IN$,
which gives
\begin{align}\nonumber\\[-2.7em]\label{Max-small-stress}
\varrho\Dt v=
  \sigma_x\ \ \ \text{ and }\ \ \ \frac{\Dt{\sigma}}{C}+\frac\sigma{\Kv}=v_x\,.
\end{align}
This leads to the dispersive wave equation
\begin{align}\label{telegraph-eq}
\frac1{C}\DDt\sigma+\frac1D\Dt\sigma-\frac1\varrho\sigma_{xx}=0 \,.
\end{align}
Alternatively, we can eliminate $\sigma$ which leads to
the same dispersive equation but in terms of $u$ instead of $\sigma$.
 Wave equations with such types of weak dumping are called telegraph
equations, exhibiting a hyperbolic rather than a parabolic character and
allowing for transmission of ultra-high frequency (i.e., ultra short length)
waves, cf.\ Figure-\ref{KV-dispersion-}-dash lines.  
Using \eq{telegraph-eq} in terms of $u$ with the ansatz \eq{dispersion-ansatz},
we obtain $w^2/{C}-{\rm i}w/D-1/(\varrho\lambda^2)=0$. In terms of the
real-valued coefficients $\omega$ and $\gamma$, it gives
\begin{align}\label{KK-relation+complex}
\frac{\omega^2{-}\gamma^2}{C}+{\rm i}\frac{2\omega\gamma}{C}
-{\rm i}\frac\omega\Kv+\frac\gamma\Kv-\frac1{\varrho\lambda^2}=0\,,
\end{align}
from which we obtain the Kramers-Kronig relations here as
\begin{align}\label{KK-relation+}
&\frac1{C}(\omega^2-\gamma^2)+\frac1D\gamma-
\frac1{\varrho\lambda^2}=0
\ \ \ \ \text{ and }\ \ \ \
\frac2{C}\gamma=\frac1D,
\end{align}
Now the latter equation tells us that the attenuation 
coefficient $\gamma={C}/(2D)$ is independent of frequency, which is related to 
the low-attenuation (hyperbolic) character of Maxwell materials even under
high-frequency waves, in contrast to the parabolic rheologies as the
Kelvin-Voigt one. The former equation in \eq{KK-relation+}
yields $\omega^2={C}/(\varrho\lambda^2)
+\gamma^2-{C}\gamma/\Kv={C}/(\varrho\lambda^2)-\gamma^2
={C}/(\varrho\lambda^2)-{C}^2/(4D^2)$. Realizing that the speed 
of wave is $v=\omega\lambda$, we obtain 
\begin{align}\label{v=v(lambda)-Maxwell}
v=v(\lambda)=\sqrt{\frac {C}\varrho-\frac{{C}^2}{4\Kv^2}\lambda^2}
\end{align}
for $\lambda\le\lambda_\text{crit}:=2\Kv/\!\sqrt{\varrho {C}}$,
which reveals an {\it anomalous dispersion}, i.e.\ the high-frequency
waves propagate faster than low-frequency ones. It should be
noted that waves with length longer than $\lambda_\text{crit}$ cannot propagate
through such 1-dimensional Maxwellian media 
since the fluidic character of such media starts dominating for ultra-low
frequency waves. Conversely, ultra-short-length waves propagate with
velocity nearly as nondispersive solid $\sqrt{{C}/\varrho}$.
In fact, \eq{telegraph-eq} is a so-called telegraph equation which is known
to exhibit a {\it hyperbolic character} with only weak attenuation.
This in particular contrasts with e.g.\ Kelvin-Voigt
materials where high-frequency vibrations or waves are highly attenuated. 
Like in \eq{KV-Q-factor}, the Q-factor $1/(1-{\rm e}^{-2\lambda\gamma/v})$
now takes $\gamma={C}/(2D)$, i.e.
\begin{align}\label{Max-Q-factor}
 \text{Q-factor}\ \sim\ \frac1{1-{\rm e}^{-{C}\lambda/(\Kv v(\lambda))}}
\ \ \text{ with }\ \ v(\lambda)\ \text{ from \eq{v=v(lambda)-Maxwell}}\,.
\end{align}

\subsection{Jeffreys visco-elastodynamic model}\label{sec2.3}

More general viscoelastic rheologies may yield more general (nonmonotone)
dispersion. Let us illustrate it on the Jeffreys rheology, as in
Figure~\ref{fig1}-right. Beside homogeneous media, such model is used
for porous media (Maxwellian  polymers or rocks) filled with Newtonian fluids
\cite{SPHE20RVE}. 
It combines the additive strain decomposition in \eq{Maxewell-internal-par} with
the additive stress decomposition as in the Kelvin-Voigt model, resulting to 
the constitutive equations 
\begin{align}\label{Jeffreys-constitutive}
\sigma=\sigma_1\!+\sigma_2,\ \ \ \ u_x=e\EL\!+e\IN,\ \ \ \ 
\sigma_1=\Kv_1\Dt u_x,\ \ \ \sigma_2=\Kv_2\Dt e\IN={C}e\EL,
\end{align}
which is to be completed by the momentum equation $\varrho\Dt v=\sigma_x$.

To reveal a dispersive wave equation, we elliminate $u$ from
$\Dt \sigma/{C}+\sigma/\Kv_2=\Kv_1\DDt  u_x/{C}+(1{+}\Kv_1/\Kv_2)\Dt  u_x$,
by differentiating it in time and by substituting
$\DDDt  u_x=\Dt \sigma_{xx}/\varrho$ and $\DDt  u_x=\sigma_{xx}/\varrho$.
This gives the dispersive equation
\begin{align}\label{Jeffreys-dispers}
\frac1{C}\DDt\sigma+\frac1{\Kv_2}\Dt\sigma-\frac{\Kv_1}{\varrho {C}}\Dt\sigma_{xx}
-\bigg(1{+}\frac{\Kv_1}{\Kv_2}\bigg)\frac1\varrho\sigma_{xx}=0\,.
\end{align}
Exploiting the ansatz \eq{dispersion-ansatz} for \eq{Jeffreys-dispers}
written for $u$ instead of $u$,
we obtain $w^2/{C}-{\rm i}w/\Kv_2-{\rm i}\Kv_1w/(\varrho {C}\lambda^2)
-(1{+}\Kv_1/\Kv_2)/(\varrho\lambda^2)=0$. 
In terms of the real-valued coefficients $\omega$ and $\gamma$, it gives
$(\omega^2{-}\gamma^2+2{\rm i}\omega\gamma)/{C}+(\gamma-{\rm i}\omega)/\Kv_2
+(\gamma-{\rm i}\omega)\Kv_1/(\varrho {C}\lambda^2)
-(1{+}\Kv_1/\Kv_2)/(\varrho\lambda^2)=0$.
The Kramers-Kronig relations are here
\begin{align}\label{Jeffreys-KK-relation}
\frac{\omega^2{-}\gamma^2}{C}+\frac\gamma{\Kv_2}
+\frac{\gamma\Kv_1}{\varrho {C}\lambda^2}
-\frac{\Kv_1{+}\Kv_2}{\varrho\Kv_2\lambda^2}
=0\ \ \ \text{ and }\ \ \
\frac{2\gamma}{C}-\frac1{\Kv_2}-\frac{\Kv_1}{\varrho {C}\lambda^2}=0\,.
\end{align}
From this we obtain the expression for $v=\omega\lambda$ as
\begin{align}\label{v=v(lambda)-Jeffreys}
v(\lambda)=\!\sqrt{
\bigg(1{+}\frac{\Kv_1}{\Kv_2}\bigg)\frac {C}\varrho+\gamma^2\lambda^2
-\frac{\gamma\Kv_1}\varrho-
\frac{\gamma {C}}{\Kv_2}\lambda^2}\ \ \text{ with }\
\gamma=\frac {C}{2\Kv_2\!}+\frac{\Kv_1}{\!2\varrho\lambda^2}
\end{align}
and 
\begin{align}\label{Jeffreys-Q-factor}
\text{Q-factor }\sim\ \frac1{1-{\rm e}^{-2\lambda\gamma/v(\lambda)}}\ \
\text{ with $v(\lambda)$ and
$\gamma=\gamma(\lambda)$ from \eq{v=v(lambda)-Jeffreys}}\,. 
\end{align}
Note that, for $\Kv_1=0$, the Jeffreys model turns into the Maxwell rheology,
i.e.\ the dispersive wave equation \eq{Jeffreys-dispers} naturally turns
\eq{telegraph-eq} with $\Kv=\Kv_2$ and thus also \eq{v=v(lambda)-Jeffreys}
turns into \eq{v=v(lambda)-Maxwell}. Conversely, for $\Kv_2=\infty$, the
Jeffreys model turns into the Kelvin-Voigt rheology,
i.e.\ the dispersive wave equation \eq{Jeffreys-dispers} naturally turns 
\eq{dispersion-} with $\Kv=\Kv_1$ and thus also \eq{v=v(lambda)-Jeffreys} turns
into \eq{v=v(lambda)}.

\begin{figure}[ht]\hspace*{5em}
\begin{tikzpicture}  
    \begin{axis}[width=6.5 cm,height=6.2 cm,
        xmin=-0.3,xmax=15,ymin=-0.1,ymax=1.7,  
        clip=true,  
        axis lines=center,  
        grid = major,  
        ytick={0, 0.5, 1, 1.5},  
        xtick={0, 2,...,15},  
         xlabel=$\dfrac1k$, ylabel={$v=v(1/k)$\hspace*{-2em}},  
    every axis y label/.style={at=(current axis.above origin),anchor=south},  
    every axis x label/.style={at=(current axis.right of origin),anchor=west},  
      ]  
\addplot[very thick,domain=0:15,samples=60] {sqrt(max(0,1-(3/(2*x))^2))};
\addplot[very thick,domain=0:15,samples=50,densely dashed]{sqrt(max(0,1-1*x^2/(4*7^2)))};
\addplot[very thick,domain=0:15,samples=50,dotted]
{sqrt(max(0,(1+3/7)*1+(1/(2*7)+3/(2*x^2))^2*x^2-(1/(2*7)+3/(2*x^2))*(3+(x^2/7))))};
 \legend{$\mbox{\scriptsize {\sl Kelvin-Voigt} \eqref{v=v(lambda)}}$,
      {\scriptsize {\sl Maxwell} \eqref{v=v(lambda)-Maxwell}\ \ } ,
  {\scriptsize {\sl Jeffreys} \eqref{v=v(lambda)-Jeffreys}\ \ } ,
            };
            \end{axis}  
  \end{tikzpicture}\hspace*{.5em}
\begin{tikzpicture}  
    \begin{axis}[width=6.5 cm,height=6. cm,
      xmin=-0.1,xmax=15,ymin=.99,ymax=4.3,  
        clip=true,  
        axis lines=center,  
        grid = major,  
        ytick={1, 2, 3, 4},  
        xtick={0, 2,...,15},  
       xlabel=$\dfrac1k$, ylabel={$Q=Q(1/k)$\hspace*{-2em}},  
   every axis y label/.style={at=(current axis.above origin),anchor=south},  
   every axis x label/.style={at=(current axis.right of origin),anchor=west},  
      ]  
\addplot[very thick,domain=0.1:14.5,samples=50] {1/(1-exp(-3/(x*sqrt(max(0.0001,1-(3/(2*x))^2)))))};
\addplot[very thick,domain=0:14.5,samples=50,densely dashed]{1/(1-exp(-1*x/(7*sqrt(max(0.0001,1-1*x^2/(4*7^2))))))};
\addplot[very thick,domain=0:15,samples=50,dotted] {1/(1-exp(-2*x*(1/(2*7)+3/(2*x^2))/sqrt(max(0.001,(1+3/7)*1+(1/(2*7)+3/(2*x^2))^2*x^2-(1/(2*7)+3/(2*x^2))*(3+(x^2/7))))))};
 \legend{{\scriptsize {\sl Kelvin-Voigt} \eqref{KV-Q-factor}},
      {\scriptsize {\sl Maxwell} \eqref{Max-Q-factor}\ \ } ,
      {\scriptsize {\sl Jeffreys} \eqref{Jeffreys-Q-factor}\ \ } ,
             };
            \end{axis}  
  \end{tikzpicture}\hspace*{-1em}
\caption{
{\sl Dependence of the  velocity (left) and the Q-factor (right)
  on the  angular  wavelength $\lambda=1/k$ illustrating the normal,
 the anomalous,
and the general dispersion for the Kelvin-Voigt (with $\Kv=3$), Maxwell
(with $\Kv=7$), and Jeffreys (with $\Kv_1=3$ and $\Kv_2=7$)
rheologies, respectively, for ${C}$=1 and $\varrho=1$. 
}
}
\label{KV-dispersion-}
\end{figure}
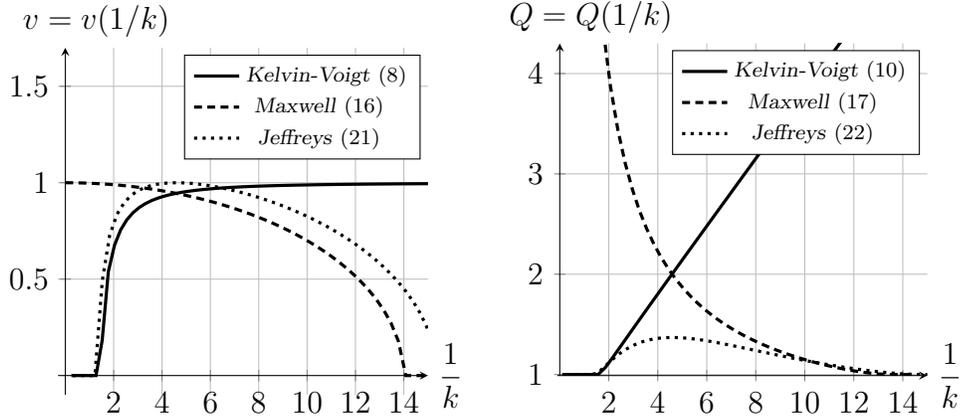

For illustration, a comparison of all these three rheological models is 
in Fig.~\ref{KV-dispersion-}. Let us note that the velocity here never
exceed the wave speed $\sqrt{{C}/\varrho}$
of the fully inviscid nondispersive model \eq{wave-eq}
which was normalized to 1 in Fig.~\ref{KV-dispersion-}  by
considering ${C}=1$ and $\varrho=1$. The viscosity moduli (physically
in Pa\,s) are also without specific units, so that Fig.~\ref{KV-dispersion-}
as well as all the following figures are to be understood as qualitative pictures
only.  Illustration of models which have capacity to exceed this velocity
in some frequency range will be done in the following section~\ref{sec3}.

\vspace*{-1em}

\begin{remark}[{\sl Standard solid: Poynting-Thomson-Zener rheology}.]\label{rem-Zener}\upshape
There is only one 3-element rheology presented in Figure~\ref{fig1}.
The other possibility of the 3-element rheology arises from the
Maxwell rheology arranged parallelly with a  Hooke elastic element or
(alternatively and, in small strains, equivalently) as Kelvin-Voigt model
arranged in series with a  Hooke  elastic element. This is called
a Zener or a Poynting-Thomson standard solid, respectively.
The explicit form of the dispersion and attenuation calculated
via the ansatz \eq{dispersion-ansatz} involves a 3rd-order
algebraic equation and leads to a complicated formula; viz
\cite[Remark~6.5.6]{KruRou19MMCM} or, in terms of phase velocity, 
\cite[Sect.2.4.3]{Carc15WFRM}.
Moreover, the standard-solid (like the Maxwell) rheology has
a hyperbolic character (manifested by easy propagation of waves with
arbitrarily high frequencies) and the analytically rigorous
large-strain variant is troublesome, as partly mentioned in
Section~\ref{sec4} below.
\end{remark}

\section{Various gradient enhancements}\label{sec3}

There are many options how to enhance the above simple-material
models by some spatial gradients. Such extensions are, at least in
some cases, related with the concept of the so-called {\it non-simple}
or {\it multipolar materials}. 
\begin{figure}[ht]
\begin{center}
\psfrag{Kelvin-Voigt-mix}{\begin{minipage}[t]{11em}\tiny Kelvin-Voigt +\\[-.2em]\hspace*{.0em}\tiny both gradients,\\[-.2em]\hspace*{.0em}\tiny  Sect.\ref{sec-KV-mixed}\end{minipage}}

\psfrag{Kelvin-Voigt-cons}{\begin{minipage}[t]{11em}\tiny Kelvin-Voigt +\\[-.2em]\hspace*{.0em}\tiny conservative\\[-.2em]\hspace*{.0em}\tiny gradient,\\[-.2em]\hspace*{.0em}\tiny  Sect.\ref{sec-KV-cons}\end{minipage}}

\psfrag{Kelvin-Voigt-dis}{\begin{minipage}[t]{11em}\tiny Kelvin-Voigt +\\[-.2em]\hspace*{.0em}\tiny dissipative\\[-.2em]\hspace*{.0em}\tiny gradient,\\[-.2em]\hspace*{.0em}\tiny  Sect.\ref{sec-KV-diss}\end{minipage}}

\psfrag{Maxwell-cons}{\begin{minipage}[t]{11em}\tiny Maxwell +\\[-.2em]\hspace*{.0em}\tiny conservative\\[-.2em]\hspace*{.0em}\tiny gradient,\\[-.2em]\hspace*{.0em}\tiny  Sect.\ref{sec-Max-cons}\end{minipage}}

\psfrag{Maxwell-dis}{\begin{minipage}[t]{11em}\tiny Maxwell +\\[-.2em]\hspace*{.0em}\tiny dissipative\\[-.2em]\hspace*{.0em}\tiny gradient,\\[-.2em]\hspace*{.0em}\tiny  Sect.\ref{sec-Max-diss}\end{minipage}}
\psfrag{Maxwell}{\small Maxwell}
\psfrag{Kelvin-Voigt}{\small Kelvin-Voigt}

\psfrag{Jeffreys}{\begin{minipage}[t]{11em}\tiny Jeffreys +\\[-.2em]\hspace*{.0em}\tiny dissipative\\[-.2em]\hspace*{.0em}\tiny gradients,\\[-.2em]\hspace*{.0em}\tiny  Sect.\ref{sec-Jeff-diss}\end{minipage}}

\psfrag{C}{\footnotesize ${C}$}
\psfrag{C,l}{\footnotesize ${C},\ell$}
\psfrag{C,l2}{\footnotesize ${C},\ell_2$}
\psfrag{D}{\footnotesize $\Kv$}
\psfrag{D,l}{\footnotesize $\Kv,\ell$}
\psfrag{D1}{\footnotesize $\Kv_1$}
\psfrag{D2}{\footnotesize $\Kv_2$}
\psfrag{D1}{\footnotesize $\Kv_1$}
\psfrag{l1}{\footnotesize $\ell_1$}
\psfrag{D2}{\footnotesize $\Kv_2$}
\psfrag{l2}{\footnotesize $\ell_2$}
\psfrag{ein}{\footnotesize $e\IN$}
\psfrag{eel}{\footnotesize $e\EL$}
\psfrag{s1}{\footnotesize $$}
\psfrag{s2}{\footnotesize $$}
\psfrag{e}{\footnotesize $e=u_x^{}$}
\hspace*{.0em}\includegraphics[width=40em]{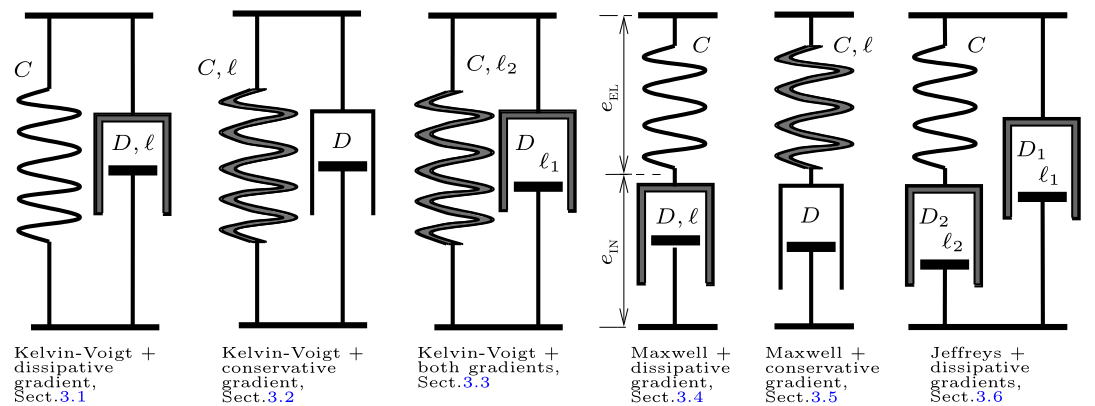}
\end{center}
\vspace*{-1em}
\caption{
  {\sl Schematic 1D-diagrammes of some gradient-enhanced  rheologies 
    from Figure~\ref{fig1}; the enhanced elements are depicted in wider lines.
    Some other gradient enhancements from Sections~\ref{sec-Brenner}  and
    \ref{sec-kinematic-grad}  do not bear such a figuration
     and thus are not included in this diagram.}
}
\label{fig2}
\end{figure}

\subsection{Dissipative-gradient Kelvin-Voigt rheology}\label{sec-KV-diss}

Let us start with a gradient enhancement of the {\it Kelvin-Voigt rheology}
by a 2nd-grade ``hyper'' viscosity with the coefficient $\ell^2{D}>0$,
i.e.\ extending \eq{dispersion-} as
\begin{align}\label{dispersion}
  \varrho \DDt u
  -\Kv \Dt u_{xx}+\ell^2{D}\Dt u_{xxxx}-{C}u_{xx}^{}=0  \,;
\end{align}
the length-scale parameter $\ell>0$ has the physical dimension meters.

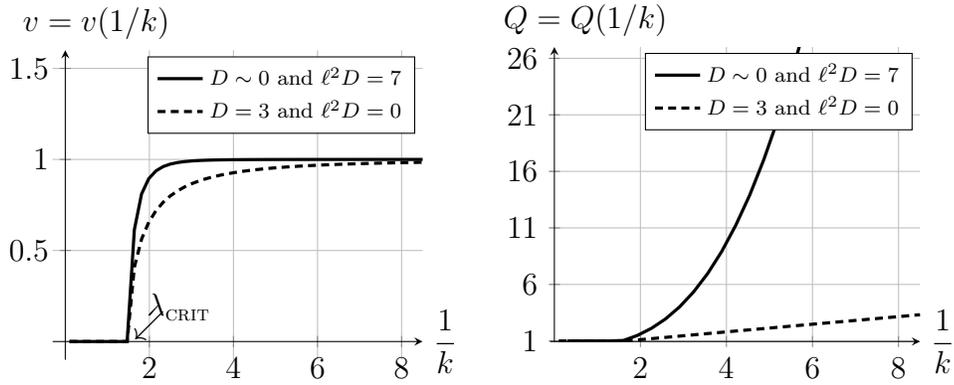
\begin{figure}[ht]\hspace*{5em}
\begin{tikzpicture}  
    \begin{axis}[width=6.5 cm,height=5.7cm,
        xmin=-0.3,xmax=8.5,ymin=-0.1,ymax=1.6,  
        clip=true,  
        axis lines=center,  
        grid = major,  
        ytick={0, 0.5, 1, 1.5},  
        xtick={0, 2,...,8},  
       xlabel=$\dfrac1k$, ylabel={$v=v(1/k)$\hspace*{-2em}},  
        every axis y label/.style={at=(current axis.above origin),anchor=south},  
        every axis x label/.style={at=(current axis.right of origin),anchor=west},  
      ]  
\addplot +[mark=none,black,very thick,domain=0.1:8.5,samples=50] {sqrt(max(0,1-(7/(2*x^3))^2))};
\addplot +[mark=none,densely dashed,black,very thick,domain=0.1:8.5,samples=50]{sqrt(max(0,1-(3*x^2+0)^2/(4*x^6)))};
\node at (axis cs:2.,0.09) {${\Large\swarrow}$};
\node at (axis cs:2.7,0.19) {$\lambda_\text{\sc crit}$};   
\legend{{\scriptsize  $D\sim 0\ \text{and}\ \ell^2{D}=7$},
 {\scriptsize  $D=3\ \text{and}\ \ell^2{D}=0$},};
           \end{axis}  
  \end{tikzpicture}\hspace*{.5em}
\begin{tikzpicture}  
    \begin{axis}[width=6.5 cm,height=5.5 cm,
        xmin=-0.1,xmax=8.5,ymin=.95,ymax=27,  
        clip=true,  
        axis lines=center,  
        grid = major,  
        ytick={1, 6, 11, 16, 21, 26},  
        xtick={0, 2,...,8},  
         xlabel=$\dfrac1k$, ylabel={$Q=Q(1/k)$\hspace*{-2em}},  
        every axis y label/.style={at=(current axis.above origin),anchor=south},  
        every axis x label/.style={at=(current axis.right of origin),anchor=west},  
      ]  
\addplot +[mark=none,black,very thick,domain=.3:6.5,samples=20] {1/(1-exp(-7/(x^3*sqrt(max(0.1,1-(7/(2*x^3))^2)))))};
\addplot +[mark=none,densely dashed,black,very thick,domain=0.1:8.5,samples=20]{1/(1-exp(-3/(x*sqrt(max(0.05,1-(3*x^2+0)^2/(4*x^6))))))};
 \legend{{\scriptsize  $D\sim 0\ \text{and}\ \ell^2{D}=7$},
 {\scriptsize  $D=3\ \text{and}\ \ell^2{D}=0$},};
            \end{axis}  
  \end{tikzpicture}\hspace*{-1em}
\caption{
{\sl A comparison of the (normal) dispersion and Q-factor due to the ``hyper''
viscosity (solid line) with the standard viscosity in the Kelvin-Voigt  viscoelastic
model as in Figure~\ref{KV-dispersion-} (dashed line).
For the same $\lambda_\text{\sc crit}$ from \eq{crit}, the dispersion due to
hyper-viscosity can be less visible than the dispersion due to usual simple
viscosity and facilitates easier wave propagation due to a higher Q-factor.
}
}
\label{KV-dispersion-comparison}
\end{figure}

Having in mind the ansatz \eq{dispersion-ansatz},
we have $\varrho\DDt  u=-\varrho w^2u$,
$\Kv\Dt  u_{xx}=-{\rm i}\Kv w u/\lambda^2$,
$\ell^2{D}\Dt  u_{xxxx}={\rm i}\ell^2{D} w u/\lambda^4$, and 
${C}u_{xx}=-{C}u/\lambda^2$, so that the one-dimensional dispersive wave equation
\eq{dispersion} yields the algebraic condition 
\begin{align}\label{KK-relation-grad}
\varrho w^2
-{\rm i}
\,\bigg(\frac\Kv{\lambda^2}{+}\frac{\ell^2{D}}{\lambda^4}\bigg)\,w
-\frac {C}{\lambda^2}=0\,.
\end{align}
When substituting $w=\omega+{\rm i}\gamma$ so that $w^2=\omega^2-\gamma^2
+2{\rm i}\omega\gamma$, we obtain the Kramers-Kronig relations as
\begin{align}\label{KK-relation-dissip-KV}
&\varrho(\omega^2-\gamma^2)=\frac {C}{\lambda^2}-
\bigg(\frac\Kv{\lambda^2}{+}\frac{\ell^2{D}}{\lambda^4}\bigg)\gamma
\ \ \ \text{ and }\ \ \
2\varrho\gamma=\frac\Kv{\lambda^2}{+}\frac{\ell^2{D}}{\lambda^4}\,.
\end{align}
From the latter equation we can read that $\gamma=
\Kv(\lambda^2+\ell^2)/(2\varrho\lambda^4)$
and then the former equation yields $(\lambda\omega)^2={C}/\varrho
+\gamma^2\lambda^2-\Kv(1{+}\ell^2/\lambda^2)\gamma/\varrho
={C}/\varrho-\gamma^2\lambda^2={C}/\varrho-\Kv^2(\lambda^2{+}\ell^2)^2/(4\varrho^2\lambda^6)$.
Realizing that the speed of waves is $v=\omega\lambda$, we obtain 
\begin{align}\nonumber\\[-2.7em]
  \label{v=v(lambda)-gradient}
v=v(\lambda)=\sqrt{\frac {C}\varrho
-\Kv^2\frac{(\lambda^2{+}\ell^2)^2}{4\varrho^2\lambda^6}}\,,
\end{align}
which gives a {\it normal dispersion} for sufficiently
long waves, namely having a length $\lambda>\lambda_\text{\sc crit}$ with the
critical wavelength $\lambda_\text{\sc crit}>0$ solving the equation
\begin{align}\label{crit}
2\sqrt{\varrho {C}}\lambda_\text{\sc crit}^3-D\lambda_\text{\sc crit}^2-\ell^2{D}=0\,,
\end{align}
cf.\ Fig.~\ref{KV-dispersion-comparison} for illustration. 
Let us recall that the adjective ``normal'' for dispersion means that waves
with longer lengths propagate faster than those with shorter lengths.
The attenuation $\gamma=\gamma(\lambda)$ is decreasing with the
wavelength $\lambda$, cf.\ the latter relation in \eq{KK-relation-dissip-KV}.
In particular, waves with the 
length $\lambda_\text{\sc crit}$ or shorter are so much attenuated that 
they cannot propagate at all.
This also reveals the dispersion/attenuation for the simple Kelvin-Voigt
model from Sect.~\ref{sec2.1} when putting $\ell^2=0$ into
\eq{v=v(lambda)-gradient} and \eq{crit}; in particular
$\lambda_\text{\sc crit}=\Kv/\!\sqrt{4\varrho {C}}$.
The other extreme case is for the inviscid situation $\Kv\to0+$ where
$\lambda_\text{\sc crit}\to0+$ so that also waves with ultra-high frequencies
can propagate.

As for the Q-factor $1/(1{-}{\rm e}^{-2\lambda\gamma/v})$
in the ``hyper'' Kelvin-Voigt model \eq{dispersion}, here 
$\gamma=\gamma(\lambda)$ determined from the latter equality in \eq{KK-relation-dissip-KV}
and $1/\omega=\lambda/v(\lambda)$  where the velocity $v(\lambda)$
is here considered from \eq{v=v(lambda)-gradient}, i.e.
\begin{align}\label{KV-hyper-visc-Q-factor}
 \text{Q-factor}\ \sim\ \frac1{1-{\rm e}^{-\Kv(\ell^2+\lambda^2)/(\varrho\lambda^3v(\lambda))}}
\ \ \text{ with }\ \ v(\lambda)\ \text{ from \eq{v=v(lambda)-gradient}}\,.
\end{align}

There is an interesting question about what is the difference between the
normal dispersion caused by standard simple viscosity and those caused by
higher-grade multipolar viscosities. In Fig.~\ref{KV-dispersion-comparison},
we can see a comparison of normal dispersion due to the conventional
viscosity $\Kv=10$ with $\ell=0$ or due to the hyperviscosity $\Kv\sim 0$ and
$\ell^2{D}=30$ with the same critical  angular  wavelength
$\lambda_\text{\sc crit}$.

\begin{remark}[{\sl Energetics for $\ell\to0$.}]\label{rem-KV-hyper-enrg}\upshape
Considering \eq{dispersion} on the one-dimensional domain
$\varOmega=[0,1]$ with zero traction boundary conditions $u_x=u_{xx}=0$
at $x=0$ and $=1$, the energetics behind \eq{dispersion} can be
seen by testing it by $\Dt u$ and integrating over $\varOmega$. This
gives
\begin{align}\label{KV-hyper-enrg}
  \frac{\d}{\d t}\int_\varOmega\hspace{-.7em}\lineunder{\frac\varrho2\big(\Dt u \big)^2\!\!+
    \frac{C}2(u_x)^2}{kinetic and stored energy}\hspace{-.7em}\d x
  +\int_\varOmega\hspace{-.2em}\lineunder{\Kv\big(\Dt u_x\big)^2\!\!
    +\ell^2\Kv\big(\Dt u_{xx}\big)^2}{dissipation rate}\hspace{-.2em}\d x=0\,.
\end{align}
The concept of nonsimple materials and, in particular, the higher-gradient
extension is often considered questionable and it is therefore interesting
to ask how it influences the energetics for small $\ell>0$.
When testing \eq{dispersion} by $\Dt u_{xx}$, we obtain
\begin{align*}
  \frac{\d}{\d t}\int_\varOmega\frac\varrho2\big(\Dt u_x\big)^2\!\!+
    \frac{C}2(u_{xx})^2\d x
  +\int_\varOmega\Kv\big(\Dt u_{xx}\big)^2\!\!
    +\ell^2\Kv\big(\Dt u_{xxxx}\big)^2\,\d x=0\,.
\end{align*}
From this, considering enough regular (smooth) initial conditions, namely
$u_{xx}\!\!\mid_{t=0}^{}$ and $\Dt u_{x}\!\!\mid_{t=0}^{}$ square-integrable on
$\varOmega$, we can see that $\int_\varOmega\Kv(\Dt u_{xx})^2\,\d x$ is
square-integrable in time. As a result, the dissipation due to
the hyper-viscosity in \eq{KV-hyper-enrg} vanishes for $\ell\to0$.
In particular, the energetics of the model for small $\ell>0$ is
not much different from the energetics of the simple Kelvin-Voigt model
\eq{dispersion-}. These energetical considerations holds also for
multidimensional situations, cf.\ also \cite[Prop.\,2]{RajRou03EDSM}.
\end{remark}

\subsection{Conservative-gradient Kelvin-Voigt rheology}\label{sec-KV-cons}

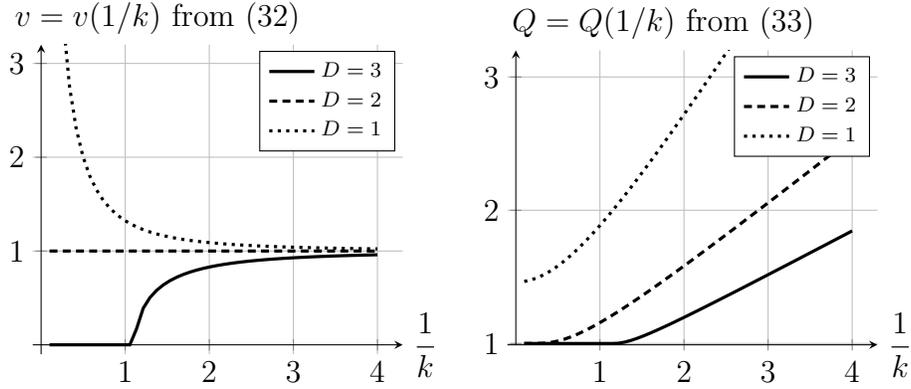
\begin{figure}[ht]\hspace*{5em}
\begin{tikzpicture}  
    \begin{axis}[width=6.5 cm,height=5.7 cm,
        xmin=-0.1,xmax=4.3,ymin=-0.1,ymax=3.2,  
        clip=true,  
        axis lines=center,  
        grid = major,  
        ytick={0, 1, ...,3},  
        xtick={0, 1,...,4},  
     xlabel=$\dfrac1k$, ylabel={$v=v(1/k)$ from \eqref{v-anomalous}\hspace*{-7.5em}},  
        every axis y label/.style={at=(current axis.above origin),anchor=south},  
        every axis x label/.style={at=(current axis.right of origin),anchor=west},  
      ]  
\addplot +[mark=none,black,very thick,domain=0.1:4,samples=50] {sqrt(max(0,1+(1*1^2-3^2/4)/x^2))};
\addplot +[mark=none,black,very thick,domain=0.1:4,samples=50,densely dashed] {sqrt(max(0,1+(1*1^2-2^2/4)/x^2))};
\addplot +[mark=none,black,very thick,domain=0.1:4,samples=50,dotted] {sqrt(max(0,1+(1*1^2-1^2/4)/x^2))};
   \legend{
      {\scriptsize  $\Kv=3$},
      {\scriptsize  $\Kv=2$},
      {\scriptsize  $\Kv=1$},
             };
            \end{axis}  
  \end{tikzpicture}\hspace*{.5em}
\begin{tikzpicture}  
    \begin{axis}[width=6.5 cm,height=5.5 cm,
        xmin=-0.1,xmax=4.3,ymin=.99,ymax=3.2,  
        clip=true,  
        axis lines=center,  
        grid = major,  
        ytick={1, 2, 3},  
        xtick={0, 1,...,4},  
        xlabel=$\dfrac1k$, ylabel={$Q=Q(1/k)$ from \eqref{v-anomalous+Q}\hspace*{-9.5em}},
        every axis y label/.style={at=(current axis.above origin),anchor=south},  
        every axis x label/.style={at=(current axis.right of origin),anchor=west},  
      ]  
\addplot +[mark=none,black,very thick,domain=0.1:4,samples=50] {1/(1-exp(-3/(x*sqrt(max(0.01,1+(1*1^2-3^2/4)/x^2)))))};
\addplot +[mark=none,black,very thick,domain=0.1:4,samples=50,densely dashed] {1/(1-exp(-2/(x*sqrt(max(0.01,1+(1*1^2-2^2/4)/x^2)))))};
\addplot +[mark=none,black,very thick,domain=0.1:4,samples=50,dotted] {1/(1-exp(-1/(x*sqrt(max(0.01,1+(1*1^2-1^2/4)/x^2)))))};
   \legend{
      {\scriptsize  $\Kv=3$},
      {\scriptsize  $\Kv=2$},
      {\scriptsize  $\Kv=1$},
             };
            \end{axis}  
  \end{tikzpicture}\hspace*{-1em}
\caption{
{\sl
Normal (solid line) and anomalous (dotted line) dispersion which both can be obtained by the
model \eq{dispersion+} as well as the nondispersive variant (dashed line);
$\ell=1$, ${C}$=1 and $\varrho=1$.}}
\label{KV-dispersion-anomalous}
\end{figure}

Another variant of dispersion by incorporating higher-order spatial
gradients into the Kelvin-Voigt rheology
is in the conservative part of the system by augmenting the
stored energy by the strain gradient as $\varphi=\frac12{C}(e^2+\ell^2e_x^2)$
where $\ell>0$ is a length-scale parameter (in meters). Sometimes, this
gradient term is interpreted as {\it capillarity}.
In the Kelvin-Voigt model \eq{dispersion-}, this leads a dispersive wave equation
\begin{align}\label{dispersion+}
  \varrho \DDt u-\Kv \Dt u_{xx}-{C}\big(u{-}\ell^2u_{xx}^{}\big)_{xx}^{}=0\,.
\end{align}
This corresponds to expansion of the stored energy $\frac12{C}u_x^2$ in
\eq{dispersion-} by the term  $\frac12\ell{C}u_{xx}^2$.
Instead of \eq{KK-relation-}, using also
$\ell^2 {C}u_{xxxx}^{}=\ell^2 {C}u/\lambda^4$, we now have
$\varrho w^2-{\rm i}\Kv w/\lambda^2-{C}/\lambda^2-\ell^2 {C}/\lambda^4=0$
so that the Kramers-Kronig relations \eq{KK-relation} now expands as
\begin{align}
\label{KK-relation+hyper}
&\varrho(\omega^2-\gamma^2)=\frac {C}{\lambda^2}-\Kv\frac\gamma{\lambda^2}+\ell^2 \frac {C}{\lambda^4}
\ \ \ \text{ and }\ \ \
2\varrho\gamma=\frac{\Kv}{\lambda^2}\,;
\end{align}
the length-scale parameter $\ell>0$ has again the physical dimension meters. 
The speed of wave $v=\omega\lambda$ is now
\begin{align}\nonumber\\[-2.7em]\label{v-anomalous}
v=v(\lambda)=
\sqrt{\frac {C}\varrho+\bigg(\ell^2{C}{-}\frac{\Kv^2}{4\varrho}\bigg)\frac1{\varrho\lambda^2}}
\,.
\end{align}
The corresponding Q-factor $1/(1{-}{\rm e}^{-2\lambda\gamma/v})$
uses $\gamma=\Kv/(2\varrho\lambda^2)$ from \eq{KK-relation+hyper}, i.e.
\begin{align}\label{v-anomalous+Q}
 \text{Q-factor}\ \sim\ \frac1{1-{\rm e}^{-\Kv/(\varrho\lambda v(\lambda))}}
\ \ \text{ with }\ \ v(\lambda)\ \text{ from \eq{v-anomalous}}\,.
\end{align}
Interestingly, for $\ell>2\sqrt{\varrho {C}}/\Kv$, the conservative
gradient facilitates the propagation of high-frequency waves, which otherwise
could not propagate in the standard Kelvin-Voigt model,
an {\it anomalous dispersion}, i.e.\
higher-frequency waves (i.e.\ with longer wavelengths) propagate faster
than waves with lower frequencies. Moreover,
for the special ratio of $\Kv$ and $\ell$, namely
\begin{align}\label{KV-nondispersive}
\frac{\Kv}{\ell}=2\sqrt{\varrho {C}}\,,
\end{align}
we obtain a non-dispersive model with the velocity $v=\sqrt{{C}/\varrho}$ as
in the basic elastodynamic model $\varrho\DDt  u={C}u_{xx}$ but now with
a finite $\lambda$-dependent Q-factor.
In the opposite case $\Kv/\ell>2\sqrt{\varrho {C}}$,
we obtain the normal dispersion. No matter what is dispersion, for  
$\Kv>0$ (very) small, the Q-factor is (very) large. In particular, we can
thus devise a gradient solid/parabolic-type model (which might be advantageous
from mathematical reasons especially if there would be some nonlinearities
involved) which is nondispersive and possibly a very low attenuation (high Q-factor),
viz Figure~\ref{KV-dispersion-anomalous} (dashed line).

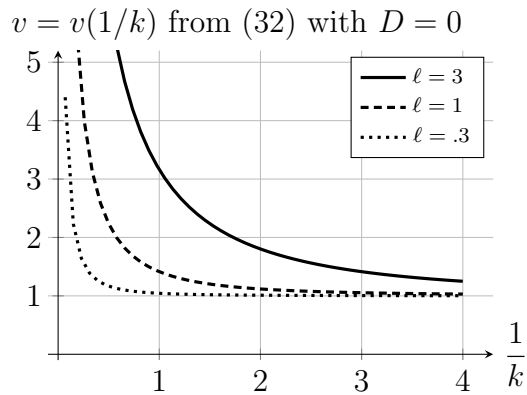
\begin{figure}[ht]\hspace*{11em}
\begin{tikzpicture}  
    \begin{axis}[width=7.5 cm,height=5.7 cm,
        xmin=-0.1,xmax=4.3,ymin=-0.1,ymax=5.2,  
        clip=true,  
        axis lines=center,  
        grid = major,  
        ytick={0, 1, ...,5},  
        xtick={0, 1,...,4},  
       xlabel=$\dfrac1k$, ylabel={$v=v(1/k)$ from \eqref{v-anomalous} with $\Kv=0$\hspace*{-11.5em}},
        every axis y label/.style={at=(current axis.above origin),anchor=south},  
        every axis x label/.style={at=(current axis.right of origin),anchor=west},  
      ]  
\addplot +[mark=none,black,very thick,domain=0.1:4,samples=50] {sqrt(max(0,1+(1*3^2)/x^2))};
\addplot +[mark=none,black,very thick,domain=0.1:4,samples=50,densely dashed] {sqrt(max(0,1+(1*1^2)/x^2))};
\addplot +[mark=none,black,very thick,domain=0.07:4,samples=50,dotted] {sqrt(max(0,1+(1*.3^2)/x^2))};
   \legend{
      {\scriptsize  $\ell=3\ $},
      {\scriptsize  $\ell=1\ $},
      {\scriptsize  $\ell=.3$},
            };
            \end{axis}  
  \end{tikzpicture}\hspace*{-.5em}
\caption{
{\sl
  Anomalous dispersion of a conservative model \eq{dispersion+} with $\Kv=0$
  for various length-scale parameters $\ell$; ${C}$=1 and $\varrho=1$.
  The Q-factor is identically $+\infty$.}}
\label{Hook-dispersion-anomalous}
\end{figure}

A special situation occurs if $\Kv=0$, which makes the model
\eq{dispersion+} conservative. In particular, there is no attenuation
so the Q-factor is $+\infty$. Actually, it is the conservative
enhancement of \eq{wave-eq} leading to anomalous dispersion
illustrated in Figure~\ref{Hook-dispersion-anomalous}, like
considered in \cite[Sect.\,1.2.2]{BerVan17IVT}, 
\cite[Sec.2.2]{Jira04NTCM}, or \cite{LiWeWa19DFEW}.

\vspace*{-1em}

\begin{remark}[{\sl Energetics for $\ell\to0$.}]\label{rem-KV-hyper-enrg+}\upshape
  Considering \eq{dispersion+} on the one-dimensional domain
  as in Remark~\eq{rem-KV-hyper-enrg},
   its energetics can again be seen by testing it by $\Dt u$ and integrating
  over $\varOmega$, which gives 
\begin{align}\label{KV-hyper-enrg+}
  \frac{\d}{\d t}\int_\varOmega\hspace{-.0em}\lineunder{\frac\varrho2\big(\Dt u\big)^2\!\!+
    \frac{C}2(u_x)^2+\ell^2{C}(u_{xx})^2}{kinetic and stored energy}\hspace{-.0em}\d x
  +\int_\varOmega\hspace{-.7em}\lineunder{\Kv\big(\Dt u_x\big)^2\!\!
  }{dissipation rate}\hspace{-.7em}\d x=0\,.
\end{align}
To show the asymptotics for $\ell\to0$ like in Remark~\eq{rem-KV-hyper-enrg},
we test \eq{dispersion+} by $\Dt u_{xxxx}$. This yields
\begin{align*}
  \frac{\d}{\d t}\int_\varOmega\frac\varrho2\big(\Dt u_{xx}\big)^2\!\!+
    \frac{C}2(u_{xxx})^2\!+\ell^2\frac{C}2(u_{xxxx})^2\d x
  +\int_\varOmega\Kv\big(\Dt u_{xxx}\big)^2\,\d x=0\,.
\end{align*}
From this, considering the initial conditions  $u_{xxx}\!\!\mid_{t=0}^{}$
and $\Dt u_{x}\!\!\mid_{t=0}^{}$ square-integrable on
$\varOmega$, we can see that $\int_\varOmega{C}(\Dt u_{xx})^2\,\d x$ is
uniformly bounded in time. As a result, the higher-gradient stored energy
term in \eq{KV-hyper-enrg+} integrated in time vanishes for $\ell\to0$.
In particular, the energetics of the model for small $\ell>0$ is
not much different from the energetics of the simple Kelvin-Voigt model
\eq{dispersion-}. These energetical considerations holds also for
multidimensional situations.
\end{remark}

\subsection{Mixed-gradient Kelvin-Voigt rheology}\label{sec-KV-mixed}

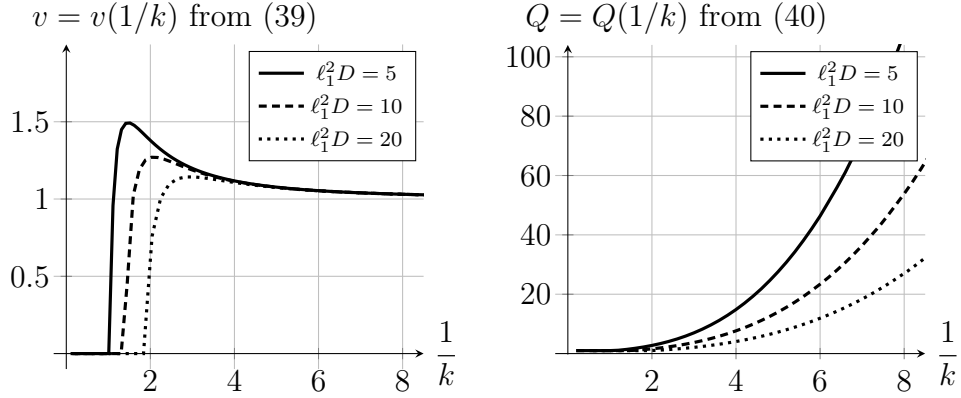
\begin{figure}[ht]\hspace*{5em}
\begin{tikzpicture}  
    \begin{axis}[width=6.5 cm,height=5.9 cm,
        xmin=-0.3,xmax=8.5,ymin=-0.1,ymax=2,  
        clip=true,  
        axis lines=center,  
        grid = major,  
        ytick={0, 0.5, 1, 1.5},  
        xtick={0, 2,...,8},  
   xlabel=$\dfrac1k$, ylabel={$v=v(1/k)$ from \eqref{v-anomalous++}\hspace*{-7em}},
        every axis y label/.style={at=(current axis.above origin),anchor=south},  
        every axis x label/.style={at=(current axis.right of origin),anchor=west},  
      ]  
\addplot[very thick,domain=0:10,samples=100]{sqrt(max(0,1*(1+(2/x)^2)-(5/x^3)^2/4))};
\addplot[very thick,domain=0:10,samples=70,densely dashed]{sqrt(max(0,1*(1+(2/x)^2)-(10/x^3)^2/4))};
\addplot[very thick,domain=0:10,samples=50,dotted]{sqrt(max(0,1*(1+(2/x)^2)-(20/x^3)^2/4))};
 \legend{{\scriptsize $\ell_1^2{D}=5$},
      {\scriptsize $\ell_1^2{D}=10$},
       {\scriptsize $\ell_1^2{D}=20$},
            };
            \end{axis}  
  \end{tikzpicture}\hspace*{.5em}
\begin{tikzpicture}  
    \begin{axis}[width=6.5 cm,height=5.7 cm,
        xmin=-0.3,xmax=8.5,ymin=-0.3,ymax=104,  
        clip=true,  
        axis lines=center,  
        grid = major,  
        ytick={0, 20, ...,100},  
        xtick={0, 2,...,8},  
        xlabel=$\dfrac1k$, ylabel={$Q=Q(1/k)$ from \eq{v-anomalous++Q}\hspace*{-7em}},
        every axis y label/.style={at=(current axis.above origin),anchor=south},  
        every axis x label/.style={at=(current axis.right of origin),anchor=west},  
      ]  
\addplot[very thick,domain=0.2:10,samples=50] {1/(1-exp(-5/(x^3*sqrt(max(0.1,1*(1+(2/x)^2)-(5/x^3)^2/4)))))};
\addplot[very thick,domain=0.2:10,samples=50,densely dashed]{1/(1-exp(-10/(x^3*sqrt(max(0.1,1*(1+(2/x)^2)-(10/x^3)^2/4)))))};
\addplot[very thick,domain=0.6:10,samples=50,dotted]{1/(1-exp(-20/(x^3*sqrt(max(0.1,1*(1+(2/x)^2)-(20/x^3)^2/4)))))};
\legend{{\scriptsize  $\ell_1^2{D}=5$},
      {\scriptsize $\ell_1^2{D}=10$},
       {\scriptsize $\ell_1^2{D}=20$},
            };
            \end{axis}  
  \end{tikzpicture}\hspace*{-1em}
\caption{
{\sl A nonmonotone dispersion of the wave velocity (left) due to \eq{v-anomalous++}
and the Q-factor (right) due to \eq{v-anomalous++Q} 
in dependence on the  angular  wavelength $\lambda=1/k$;
${C}=1$, $\varrho=1$, $\Kv\sim 0$, and $\ell_2=2$.}}
\label{KV-dispersion-mixed}
\end{figure}

Moreover, \eq{dispersion} and \eq{dispersion+} can be combined together
to get dispersion depending nonmonotonically on wavelength
by combining normal-anomalous dispersion, modelled here by gradients
in nonconservative and conservative parts. Considering now two length-scale
parameters $\ell_1>0$ and $\ell_2>0$, we have in mind the dispersive wave equation
\begin{align}\label{dispersion+-}
\varrho\DDt u-\Kv\Dt u_{xx}+\ell_1^2{D}\Dt u_{xxxx}-{C}\big(u{-}\ell_2^2u_{xx}^{}\big)_{xx}^{}=0\,.
\end{align}
The combination of \eq{KK-relation-grad} and \eq{KK-relation+complex} yields
the algebraic condition
\begin{align}\label{KK-relation+-}
\varrho w^2-{\rm i}
\,\bigg(\frac\Kv{\lambda^2}{+}\frac{\ell_1^2{D}}{\lambda^4}\bigg)\,w
-\bigg(1{+}\frac{\ell_2^2}{\lambda^2}\bigg)\frac {C}{\lambda^2}=0\,.
\end{align}
When substituting $w=\omega+{\rm i}\gamma$ so that $w^2=\omega^2-\gamma^2
+2{\rm i}\omega\gamma$, we obtain  the Kramers-Kronig relations
\begin{align}\label{KK-relation-combined}
&\varrho(\omega^2-\gamma^2)=\frac {C}{\lambda^2}-
\bigg(\frac\Kv{\lambda^2}+\frac{\ell_1^2{D}}{\lambda^4}\bigg)\gamma+\ell_2^2 \frac {C}{\lambda^4}
\ \ \ \text{ and }\ \ \
2\varrho\gamma=\frac{\Kv}{\lambda^2}{+}\frac{\ell_1^2{D}}{\lambda^4}\,.
\end{align}
The speed of wave $v=\omega\lambda$ as a function of the  angular  wavelength
is now
\begin{align}\label{v-anomalous++}
v=v(\lambda)=\sqrt{\frac {C}\varrho\bigg(1+\frac{\ell_2^2}{\lambda^2}\bigg)
-\frac{\Kv^2}{4\varrho^2\lambda^2}\bigg(1+\frac{\ell_1^2}{\lambda^2}\bigg)^2}\,.
\end{align}
The Q-factor $1/(1{-}{\rm e}^{-2\lambda\gamma/v})$ corresponding to this model
uses $\gamma=\gamma(\lambda)=(\Kv/\lambda^2+\ell_1^2{D}/\lambda^4)/(2\varrho)$ and $v=v(\lambda)$
from \eq{v-anomalous++}, i.e.
\begin{align}\label{v-anomalous++Q}
\text{Q-factor}\ \sim\ \frac1{1-{\rm e}^{-D(\ell_1^2+\lambda^2)/(\varrho\lambda^3v(\lambda))}}
\ \ \text{ with }\ \ v(\lambda)\ \text{ from \eq{v-anomalous++}}\,.
\end{align}
The combination of the normal and the anomalous dispersion in the
formulas \eq{v=v(lambda)-gradient} and \eq{v-anomalous} allows for
devising a quite general nonmonotonic dispersion, as indicated in
Figure\,\ref{KV-dispersion-mixed}.

\subsection{Maxwell rheology with dissipative gradient}\label{sec-Max-diss}

Besides the Kelvin-Voigt model, the gradient enhancement can be made also
for other rheologies. For the {\it Maxwell rheology},
it is important to start with the formulation by employing internal
variable \eq{Maxewell-internal-par}. There are essentially
two options: a dissipative-gradient extension of the flow rule
$\Kv\Dt e\IN=\sigma$ as
$\Kv\Dt e\IN=\sigma+\ell^2\Kv(\Dt e\IN)_{xx}$
or a conservative-gradient extension of the  Hooke 
law ${C}e\EL=\sigma$ as ${C}e\EL=\sigma+\ell^2{C}(\Dt e\IN)_{xx}$
with some length-scale parameter $\ell>0$ again of the physical dimension
meters.

The former option of the dissipative gradient, written as
$\Kv\pi=\sigma+
\ell^2\Kv\pi_{xx}$ when abbreviating by $\pi:=\Dt e\IN$
the {\it rate of the inelastic creep strain} $e\IN$, leads to
\begin{align}\label{hyper-Maxwell-1D}
\varrho\DDt u=\sigma_x,\ \ \ \ \ \ \Kv\big(\pi{-}\ell^2\pi_{xx}\big)
=\sigma,\ \ \text{ and }\ \ \Dt\sigma={C}\big(\Dt u_x-\pi\big)\,.
\end{align}
Differentiating the last equation gives 
$\Dt \sigma_{xx}={C}(\Dt  u_{xxx}-\pi_{xx})$. Summing them with the weights
$\Kv$ and $-\ell^2\Kv$ allows for the elimination of the internal rate variable
$\pi$ when using the second equation in \eq{hyper-Maxwell-1D}, namely
$\Kv(\Dt \sigma-\ell^2\Dt \sigma_{xx})=\Kv {C}(\Dt  u_x-\ell^2\Dt  u_{xxx})-{C}\sigma$.
Differentiating in time and using the first equation in \eq{hyper-Maxwell-1D}
allows for elimination of $u$, yielding the dispersive wave equation
\begin{align}\label{hyper-Maxwell-1D-dispers}
  \frac1{C}
  \big(\DDt\sigma{-}\ell^2\DDt\sigma_{xx}\big)
+\frac1\Kv\Dt\sigma-\frac1\varrho\big(\sigma{-}\ell^2\sigma_{xx}^{}\big)_{xx}^{}=0\,.
\end{align}
Alternatively, we can eliminate $\sigma$:
using $\varrho\DDt u=\sigma_x$ for
$\Kv(\Dt \sigma_x-\ell^2\Dt \sigma_{xxx})=\Kv {C}(\Dt u_{xx}-\ell^2\Dt u_{xxxx})-{C}\sigma_x$
gives
$\varrho\Kv(\DDDt u-\ell^2\DDDt u_{xx})=\Kv{C}(\Dt u_{xx}-\ell^2\Dt u_{xxxx})-\varrho{C}\DDt u$.
When integrated in time, it leads to the dispersive equation
\eq{hyper-Maxwell-1D-dispers} written in terms of $u$ instead of $\sigma$.
Interestingly, the gradient term $\ell^2\pi_{xx}$ in \eq{hyper-Maxwell-1D}
is manifested in the dispersive wave equation \eq{hyper-Maxwell-1D-dispers}
written for $u$ as an inertia-gradient term $\varrho\ell^2\DDt  u_{xx}$,
cf.\ Remark~\ref{rem-microinert} below.

Exploiting the ansatz \eq{dispersion-ansatz} now for
\eq{hyper-Maxwell-1D-dispers}, we obtain
\begin{align*}
\frac{w^2}{{C}}\bigg(1{+}\frac{\ell^2}{\lambda^2}\bigg)
-{\rm i}\frac{w}{\Kv}-\frac1{\varrho\lambda^2}\bigg(1{+}\frac{\ell^2}{\lambda^2}\bigg)=0
\end{align*}
so that, realizing $w=\omega+{\rm i}\gamma$, we expand \eq{KK-relation+complex} as
\begin{align*}
\frac{\omega^2{-}\gamma^2}{C}\bigg(1{+}\frac{\ell^2}{\lambda^2}\bigg)
+{\rm i}\frac{2\omega\gamma}{{C}}\bigg(1{+}\frac{\ell^2}{\lambda^2}\bigg)
-{\rm i}\frac\omega\Kv+\frac\gamma\Kv
-\frac1{\varrho\lambda^2}\bigg(1{+}\frac{\ell^2}{\lambda^2}\bigg)=0\,.
\end{align*}
The {Kramers-Kronig relations \eq{KK-relation+} then expand as here as
\begin{align*}
&\frac1{C}(\omega^2-\gamma^2)+\frac{\gamma\lambda^2}{\Kv(\ell^2{+}\lambda^2)}
-\frac1{\varrho\lambda^2}=0
\ \ \ \ \text{ and }\ \ \ \
\frac{2\gamma}{{C}}\bigg(1{+}\frac{\ell^2}{\lambda^2}\bigg)=\frac1D\,.
\end{align*}
Thus \eq{v=v(lambda)-Maxwell} modifies as
\begin{align}\nonumber\\[-2.7em]\label{v=v(lambda)-Maxwell-grad}
v=v(\lambda)=\sqrt{\,\frac {C}\varrho-
\frac{{C}^2\lambda^6}{4\Kv^2(\lambda^2{+}\ell^2)^2}}
\end{align}
while the Q-factor is again from \eq{Max-Q-factor} but involving now \eq{v=v(lambda)-Maxwell-grad}
instead of \eq{v=v(lambda)-Maxwell}. Cf.\ Figure~\ref{Max-dispersion-grad} which
illustrates that the gradient term in \eq{hyper-Maxwell-1D} facilitates especially
the propagation of longer-length waves (as far as less dispersion and higher Q-factor
concerns) in comparison with the simple Maxwell rheology. Noteworthy, it is
an example how gradient extension of the dissipative part can lead to less
dissipation. 
\begin{figure}[ht]\hspace*{5em}
\begin{tikzpicture}  
    \begin{axis}[width=6.5 cm,height=5.7 cm,
        xmin=-0.3,xmax=22.9,ymin=-0.05,ymax=1.6,  
        clip=true,  
        axis lines=center,  
        grid = major,  
        ytick={0, 0.5, 1, 1.5},  
        xtick={0, 4,...,20},  
        xlabel=$\dfrac1k$, ylabel={$v=v(1/k)$ from \eq{v=v(lambda)-Maxwell-grad}\hspace*{-8em}},
    every axis y label/.style={at=(current axis.above origin),anchor=south},  
    every axis x label/.style={at=(current axis.right of origin),anchor=west},  
      ]  
\addplot[very thick,domain=0:22,samples=50]{sqrt(max(0,1-1*x^6/(4*(x^2+20^2)^2*7^2)))};
\addplot[very thick,domain=0:22,samples=50,densely dashed]{sqrt(max(0,1-1*x^6/(4*(x^2+10^2)^2*7^2)))};
\addplot[very thick,domain=0:22,samples=50,dotted]{sqrt(max(0,1-1*x^2/(4*7^2)))};
 \legend{{\scriptsize $\ell=20$},
      {\scriptsize $\ell=10$} ,
      {\scriptsize $\ell=0$\hspace*{.4em}} ,
             };
            \end{axis}  
  \end{tikzpicture}\hspace*{.5em}
\begin{tikzpicture}  
    \begin{axis}[width=6.5 cm,height=5.5 cm,
      xmin=-0.1,xmax=22.9,ymin=.99,ymax=12,  
        clip=true,  
        axis lines=center,  
        grid = major,  
        ytick={1, 3, 5, 7, 9, 11},  
        xtick={0, 4,...,20},  
       xlabel=$\dfrac1k$, ylabel={$Q=Q(1/k)$\hspace*{-2em}},
   every axis y label/.style={at=(current axis.above origin),anchor=south},  
   every axis x label/.style={at=(current axis.right of origin),anchor=west},  
      ]  
\addplot[very thick,domain=5:22.5,samples=50]{1/(1-exp(-1*x^3/(7*(x^2+20^2)*sqrt(max(0.09,1-1*x^6/(4*(x^2+20^2)^2*7^2))))))};
\addplot[very thick,domain=4:22.5,samples=50,densely dashed]{1/(1-exp(-1*x^3/(7*(x^2+10^2)*sqrt(max(0.09,1-1*x^6/(4*(x^2+10^2)^2*7^2))))))};
\addplot[very thick,domain=.2:22.5,samples=50,dotted]{1/(1-exp(-1*x/(7*sqrt(max(0.0001,1-1*x^2/(4*7^2))))))};
\legend{{\scriptsize $\ell=20$\hspace*{3.em}},
      {\scriptsize $\ell=10$\hspace*{3.em}} ,
      {\scriptsize $\ell\,{=}\,0$ {\smaller(Maxwell)}\hspace*{-.4em}} ,
           };
            \end{axis}  
  \end{tikzpicture}\hspace*{-1em}
\caption{
{\sl Dependence of the velocity (left) and the Q-factor (right)
  on the  angular  wavelength $\lambda=1/k$ of the Maxwell model for ${C}=1$, $\varrho=1$,
  and $\Kv=7$. It illustrates the (expected) influence of adding the hyper-viscosity to decrease
the dispersion and to increase the Q-factor.
For $\ell=0$, cf.\ the Maxwell model from Figure~\ref{KV-dispersion-}-dashed line.
}
}
\label{Max-dispersion-grad}
\end{figure}
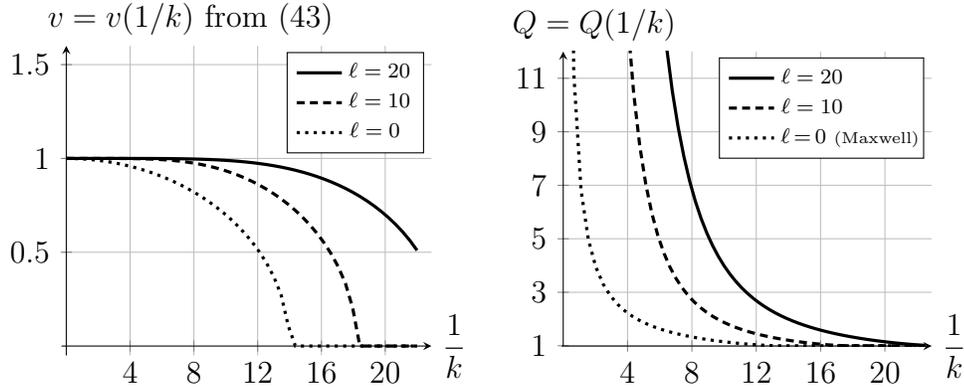

When reducing $\Kv\pi-\Kv\ell^2\pi_{xx}=\sigma$ in \eq{hyper-Maxwell-1D}
to $-H\pi_{xx}=\sigma$ with $H=\Kv\ell^2$ the hyperviscosity coefficient,
\eq{v=v(lambda)-Maxwell-grad} turns into 
\begin{align}\label{v=v(lambda)-Maxwell-grad+}
v=v(\lambda)=\sqrt{\,\frac {C}\varrho-\frac{{C}^2\lambda^6}{4H^2}}\,.
\end{align}
The Q-factor \eq{def-Q-factor} is now $1/(1-{\rm e}^{\lambda^3C/(2Hv(\lambda))})$ with
$v(\lambda)$ from \eq{v=v(lambda)-Maxwell-grad+}.
It is interesting to compare the standard viscosity in the Maxwell model with this
``hyper Maxwell'' model. When choosing $H$ so that the critical  angular 
wavelength in
\eq{v=v(lambda)-Maxwell-grad} $\lambda_\text{\sc crit}^{}=\sqrt[6]{4H^2\!/\varrho{C}}$ 
is the same as the critical  angular  wavelength $\lambda_\text{\sc crit}^{}=2\Kv/\sqrt{\varrho{C}}$ in
\eq{v=v(lambda)-Maxwell}, we can relevantly compare both models, cf.\ 
Fig.~\ref{Max-dispersion-grad+}. Similarly as in Figure~\ref{KV-dispersion-comparison}, the hyper-viscosity
facilitates the propagation of waves as far as less dispersion and higher Q-factor.
\begin{figure}[ht]\hspace*{5em}
\begin{tikzpicture}  
    \begin{axis}[width=6.7 cm,height=5.5 cm,
        xmin=-0.3,xmax=17,ymin=-0.05,ymax=1.55,  
        clip=true,  
        axis lines=center,  
        grid = major,  
        ytick={0, 0.5, 1, 1.5},  
        xtick={0, 4,...,16},  
       xlabel=$\dfrac1k$, ylabel={$v=v(1/k)$\hspace*{-2em}},
    every axis y label/.style={at=(current axis.above origin),anchor=south},  
    every axis x label/.style={at=(current axis.right of origin),anchor=west},  
      ]  
\addplot[very thick,domain=0:16,samples=50]{sqrt(max(0,1-1*x^6/(4*1372^2)))};
\addplot[very thick,domain=0:16,samples=50,dotted]{sqrt(max(0,1-1*x^2/(4*7^2)))};
\node at (axis cs:14.7,0.09) {${\Large\swarrow}$};
\node at (axis cs:15.8,0.2) {$\lambda_\text{\sc crit}$};   
\legend{{\scriptsize $v$ from \eqref{v=v(lambda)-Maxwell-grad+}, $H=1372$},
      {\scriptsize $v$ from \eqref{v=v(lambda)-Maxwell}, $\Kv=7\ \ \ \ \ $},
            };
            \end{axis}  
  \end{tikzpicture}\hspace*{.5em}
\begin{tikzpicture}  
    \begin{axis}[width=6.5 cm,height=5.5 cm,
    xmin=-0.3,xmax=16.5,ymin=.98,ymax=7.3,  
        clip=true,  
        axis lines=center,  
        grid = major,  
        ytick={1, 2, 3, 4, 5, 6, 7},  
        xtick={0, 4,...,20},  
        xlabel=$\dfrac1k$, ylabel={$Q=Q(1/k)$\hspace*{-2em}},
   every axis y label/.style={at=(current axis.above origin),anchor=south},  
   every axis x label/.style={at=(current axis.right of origin),anchor=west},  
      ]  
\addplot[very thick,domain=7.4:15.5,samples=50] {1/(1-exp(-1*x^3/(2*1372*sqrt(max(0.01,1-1*x^6/(4*1372^2))))))};
\addplot[very thick,domain=0:16,samples=50,dotted]{1/(1-exp(-1*x/(7*sqrt(max(0.0001,1-1*x^2/(4*7^2))))))};
            \end{axis}  
  \end{tikzpicture}\hspace*{-1em}
\caption{
  {\sl Comparison of the conventional Maxwellian viscosity (dotted lines) with
    the Maxwellian hyper-viscosity (solid line) with the same  angular 
    wavelength $\lambda_\text{\sc crit}^{}=14$; again ${C}=1$ and $\varrho=1$.
    The ``hyper-Maxwell'' rheology allow for propagation of waves wih
     angular  lengths below $\lambda_\text{\sc crit}^{}$ with less
    dispersion and attenuation (i.e.\ with higher Q-factor).
}
}
\label{Max-dispersion-grad+}
\end{figure}
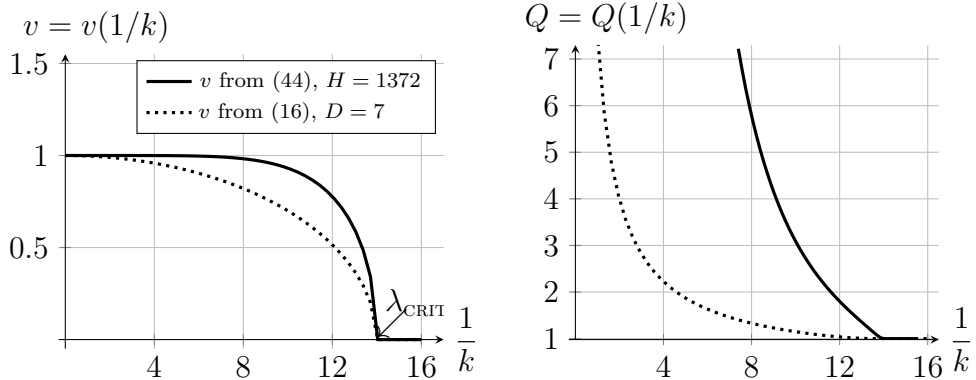

Noteworthy, for $\Kv\to\infty$, we obtain a fully conservative model with no dispersion,
i.e.\ $v=\sqrt{{C}/\varrho}$ is constant and the Q-factor is $+\infty$. This model was
devised in \cite{MetAsk02ODCG}. 

\subsection{Maxwell rheology with conservative gradient}\label{sec-Max-cons}

The later mentioned option, i.e.\ the conservative-gradient enhancement of
\eq{Maxewell-internal-par}, leads to the system
\begin{align}\label{hyper-Maxwell-1D-conserve}
\varrho\DDt u=\sigma_x,\ \ \ \ \ \Kv\pi=\sigma,\ \ \text{ and }\ \
\Dt\sigma={C}\big(\Dt u_x {-}\pi\big)-\ell^2{C}\big(\Dt u_x{-}\pi\big)_{xx}.
\end{align}
By eliminating the creep rate $\pi$ and the stress $\sigma$, we obtain
the dispersive equation
\begin{align}
  \label{telegraph-eq-alt-grad}
\frac1{{C}}\DDt u+\frac{1}{\Kv}\Dt u-\frac1\varrho u_{xx}
=\ell^2\Big(\frac{1}{\Kv}\Dt{u}-\frac1{\varrho}u_{xx}\Big)_{xx}\,.
\end{align}
Note that, for $\ell=0$, \eq{telegraph-eq-alt-grad} naturally 
turns into the telegraph equation \eq{telegraph-eq} written for $u$.
For $\ell>0$, \eq{KK-relation+complex} expands to
\begin{align*}
\frac{\omega^2{-}\gamma^2}{C}+{\rm i}\frac{2\omega\gamma}{C}
-{\rm i}\frac\omega\Kv+\frac\gamma\Kv-\frac1{\varrho\lambda^2}
=\ell^2\Big(\frac1{\varrho\lambda^4}-\frac{\gamma}{\Kv\lambda^2}
+{\rm i}\frac\omega{\Kv\lambda^2}\Big)\,,
\end{align*}
from which we obtain the Kramers-Kronig relations \eq{KK-relation+} expanded here 
as
\begin{align*}
&\frac{\omega^2{-}\gamma^2}{{C}}+\frac1D\gamma-
\frac1{\varrho\lambda^2}=\ell^2\Big(\frac1{\varrho\lambda^4}-\frac{\gamma}{\Kv\lambda^2}\Big)
\ \ \ \text{ and }\ \ \ \frac2{C}\gamma=\frac1\Kv+\frac{\ell^2}{\Kv\lambda^2}\,.\!\!
\end{align*}
This gives
\begin{align}
v=\sqrt{\frac{{C}}\varrho+\gamma\Big(\gamma{-}\frac{{C}}{\Kv}\Big)\lambda^2
+\ell^2{C}\Big(\frac1{\varrho\lambda^2}{+}\frac\gamma\Kv\Big)}\ \ \ \text{ with }
\ \gamma=\frac{{C}}{2\Kv}+\frac{\ell^2{C}}{2\Kv\lambda^2}\,.
\label{Maxwell-gradient-conserv}\end{align}
Naturally, for $\ell=0$, it reduces to the Maxwell dispersion \eq{v=v(lambda)-Maxwell}.
This is illustrated on Figure~\ref{fig-grad-Maxwell} together with the Q-factor
$1/(1{-}{\rm e}^{-2\lambda\gamma/v})$. 
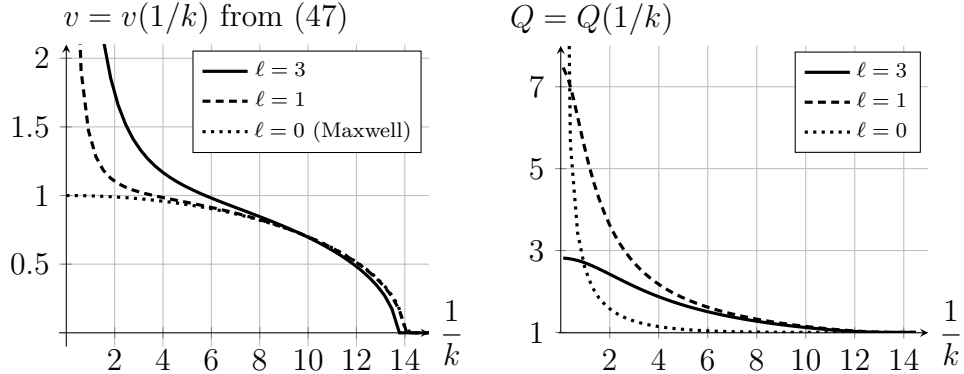
\begin{figure}[ht]\hspace*{5em}
\begin{tikzpicture}  
    \begin{axis}[width=6.5 cm,height=5.6 cm,
        xmin=-0.3,xmax=15,ymin=-0.1,ymax=2.1,  
        clip=true,  
        axis lines=center,  
        grid = major,  
        ytick={0, 0.5, 1, 1.5, 2},  
        xtick={0, 2,...,15},  
       xlabel=$\dfrac1k$, ylabel={$v=v(1/k)$ from \eqref{Maxwell-gradient-conserv}\hspace*{-9em}},
    every axis y label/.style={at=(current axis.above origin),anchor=south},  
    every axis x label/.style={at=(current axis.right of origin),anchor=west},  
      ]  
\addplot[very thick,domain=0:15,samples=50]
{sqrt(max(0,1+(1/(2*7)+3^2/(2*7*x^2))*((1/(2*7)+3^2/(2*7*x^2))-1/7)*x^2
-3^2*((1/(2*7)+3^2/(2*7*x^2))/7-1/x^2)))};
\addplot[very thick,domain=0:15,samples=50,densely dashed]
{sqrt(max(0,1+(1/(2*7)+1.0^2/(2*7*x^2))*((1/(2*7)+1.0^2/(2*7*x^2))-1/7)*x^2
-1.0^2*((1/(2*7)+1.0^2/(2*7*x^2))/7-1/x^2)))};
\addplot[very thick,domain=0:15,samples=60,dotted] {sqrt(max(0,1-1*x^2/(4*7^2)))};
\legend{{\scriptsize $\ell=3$\hspace*{4.5em}},
  {\scriptsize $\ell=1$\hspace*{4.5em}},
  {\scriptsize $\ell=0$ (Maxwell)\!},
            };
            \end{axis}  
  \end{tikzpicture}\hspace*{.5em}
\begin{tikzpicture}  
    \begin{axis}[width=6.5 cm,height=5.4 cm,
      xmin=-0.1,xmax=15,ymin=.99,ymax=8,  
        clip=true,  
        axis lines=center,  
        grid = major,  
        ytick={1, 3, 5, 7},  
        xtick={0, 2,...,15},  
        xlabel=$\dfrac1k$, ylabel={$Q=Q(1/k)$\hspace*{-2em}},
   every axis y label/.style={at=(current axis.above origin),anchor=south},  
   every axis x label/.style={at=(current axis.right of origin),anchor=west},  
      ]  
\addplot[very thick,domain=0.1:14.5,samples=70] {1/(1-exp(-2*x*(1/(2*7)+3^2/(2*7*x^2))/
sqrt(max(0.001,1+(1/(2*7)+3^2/(2*7*x^2))*((1/(2*7)+3^2/(2*7*x^2))-1/7)*x^2-3^2*((1/(2*7)+3^2/(2*7*x^2))/7-1/x^2)))))};
\addplot[very thick,domain=0.1:14.5,samples=70,densely dashed] {1/(1-exp(-2*x*(1/(2*7)+1.0^2/(2*7*x^2))/
sqrt(max(0.001,1+(1/(2*7)+1.0^2/(2*7*x^2))*((1/(2*7)+1.0^2/(2*7*x^2))-1/7)*x^2-1.0^2*((1/(2*7)+1.0^2/(2*7*x^2))/7-1/x^2)))))};
\addplot[very thick,domain=0.1:14.5,samples=50,dotted] {1/(1-exp(-1*x/(2*sqrt(max(0.0001,1-1*x^2/(4*7^2))))))};
 \legend{{\scriptsize $\ell=3$},
      {\scriptsize $\ell=1$} ,
      {\scriptsize $\ell=0$} ,
             };
            \end{axis}  
  \end{tikzpicture}\hspace*{-1em}
\caption{
{\sl Anomalous dispersion of the  velocity according \eqref{Maxwell-gradient-conserv} for $\ell=3,\,1,\,0$ (left) and the
corresponding Q-factor (right). The case $\ell=0$ is the Maxwell model from
Figure~\ref{KV-dispersion-} again ${C}$=1, $\varrho=1$, and $\Kv=7$.
}
}
\label{fig-grad-Maxwell}
\end{figure}
As expected, adding the conservative gradient into the Maxwell model as in
\eq{hyper-Maxwell-1D-conserve} facilitates propagation of ultra-high frequency
waves, i.e.\ waves with ultra-short wavelength. Notably, for $\Kv\to\infty$,
this dispersive model becomes fully conservative, $\gamma=0$, and 
\begin{align}\nonumber\\[-2.7em]
  v=\sqrt{\frac{{C}}\varrho\Big(1+\frac{\ell^2}{\lambda_2}\Big)}\,.
\end{align}
The same asymptotics towards anomalously dispersive model holds for the Kelvin-Voigt
model from Section~\ref{sec-KV-cons} with $\Kv\to0$, cf.\
Figure~\ref{Hook-dispersion-anomalous}.

\subsection{Dissipative-gradient Jeffreys rheology}\label{sec-Jeff-diss}

A dissipative gradient enhancement of the Jeffreys rheology
suggests three variants, copying either \eq{dispersion} or
\eq{hyper-Maxwell-1D}, or both. Again, we need a formulation
with the internal variable as the creep strain rate $\pi$. Thus,
\eq{Jeffreys-constitutive} extended by two gradient terms reads as
\begin{subequations}\label{Jeffreys-small-strain+++grad}\begin{align}
&\varrho\DDt u=\big(D_1\Dt u_x+{C}e+\ell_1^2\Kv_1\Dt u_{xxx}\big)_x\,,
\ \ \ \ \
\\&\Dt e=\Dt u_x-\pi\,,\ \ \text{ and }\ \ D_2(\pi-\ell_2^2\pi_{xx})={C}e
\,.
\end{align}\end{subequations}
To reveal the underlying dispersive wave equation, we make the elimination
of $\pi$ from the last and the penultimate equation in
\eq{Jeffreys-small-strain+++grad}:
\begin{align}\label{Jeffreys-small-strain+++grad+}
 \Dt e-\ell_2^2\Dt e_{xx}+\frac{C}{D_2}e=\big(\Dt u-\ell_2^2\Dt u_{xx}\big)_x
\end{align}
Then, we differentiate the first equation in \eq{Jeffreys-small-strain+++grad}
in time, which yields 
\begin{align}\label{Jeffreys$-small-strain+++grad++}
\varrho\DDDt u=\big(D_1\DDt u_x+{C}\Dt e+\ell_1^2\Kv_1\DDt u_{xxx}\big)_x\,.
\end{align}
Summing it with the first equation in \eq{Jeffreys-small-strain+++grad}
multiplied by ${C}/D_2$ and subtracting the first equation in
\eq{Jeffreys-small-strain+++grad} multiplied by $\ell_2^2$
and differentiated twice in space, we can use \eq{Jeffreys-small-strain+++grad+}
to eliminate also $e$ and obtain the dispersive wave equation in terms of the displacement
$u$, which, after integration in time, reads as
\begin{align}\nonumber
\!\varrho\big(\DDt u{-}\ell_2^2\DDt u_{xx}\big)-
{C}\big(u{-}\ell_2^2u_{xx}\big)_{xx}\!
=D_1\big(\Dt u{-}\ell_2^2\Dt u_{xx}\big)_{xx}\!
+{C}\frac{D_1}{D_2}{u_{xx}}\hspace*{4em}
\\[-.6em]
-\varrho\frac{C}{D_2}\Dt u
+\ell_1^2\Kv_1\big(\Dt u{-}\ell_2^2\Dt u_{xx}\big)_{xxxx}\!
+{C}\frac{\ell_1^2\Kv_1}{D_2}{u_{xxxx}}\,.
\label{Jeffreys-small-strain+++grad+++}\end{align}
Using the ansatz \eq{dispersion-ansatz} and abbreviating
$a(\lambda)=1+\ell_2^2/\lambda^2$, we obtain 
\begin{align*}\varrho w^2a(\lambda)
\,{-}\,{\rm i}\varrho\frac{{C}w}{D_2} 
\,{-}\,{\rm i}D_1w\frac{a(\lambda)}{\lambda^2}
\,{-}\,\frac{{C}D_1}{D_2\lambda^2}
\,{-}\,{C}\frac{a(\lambda)\!}{\lambda^2}
\,{+}\,{\rm i}w \ell_1^2\Kv_1\frac{a(\lambda)\!}{\lambda^4}
\,{+}\,\frac{{C}\ell_1^2\Kv_1\!}{D_2\lambda^4}=0\,.
\end{align*}
Reminding $w=\omega+{\rm i}\gamma$ and
$w^2=\omega^2-\gamma^2+2{\rm i}\omega\gamma$, the resulting Kramers-Kronig
re\-la\-tions are
\begin{align*}\nonumber
\varrho(\omega^2{-}\gamma^2)a(\lambda)\,{+}\,\varrho\frac{\!{C}\gamma\!}{D_2}
\,{+}\,D_1\gamma\frac{\!a(\lambda)\!}{\lambda^2}
\,{-}\,\frac{{C}D_1\!}{D_2\lambda^2\!}
\,{-}\,{C}\frac{a(\lambda)\!}{\lambda^2}
\,{-}\,\gamma \ell_1^2\Kv_1\frac{\!a(\lambda)\!}{\lambda^4}
\,{+}\,\frac{\!{C}\ell_1^2\Kv_1\!}{D_2\lambda^4}=0\,,
\\\text{ and }\ \ \ 2\varrho\gamma a(\lambda)-\varrho\frac{C}{D_2}
-D_1\frac{a(\lambda)}{\lambda^2}
+ \ell_1^2\Kv_1\frac{a(\lambda)}{\lambda^4}=0\,.
\end{align*}
Reminding $v=\omega\lambda$, we obtain the velocity $v=v(\lambda)$ as
\begin{align}\nonumber
\!\!v&=\sqrt{\frac{C}\varrho\,{+}\,\lambda^2\gamma^2(\lambda)
\,{-}\,\frac{\!C\lambda^2\gamma(\lambda)\!}{D_2a(\lambda)}
\,{-}\,\frac{\!D_1\gamma(\lambda)\!}{\varrho}
\,{+}\,\frac{{C}D_1}{\!\varrho D_2a(\lambda)\!}
\,{+}\,\frac{\!\ell_1^2\Kv_1\gamma(\lambda)\!}{\varrho\lambda^2}
\,{-}\,\frac{{C}\ell_1^2\Kv_1}{\!\varrho D_2\lambda^2 a(\lambda)\!}}
\\&\qquad\text{where }\ \ a(\lambda)=1+\frac{\ell_2^2}{\lambda^2}\ \ \text{ and }\ \ \gamma(\lambda)=\frac{C}{2D_2a(\lambda)}
+\frac{D_1}{2\varrho\lambda^2}
-\frac{\ell_1^2\Kv_1}{2\varrho\lambda^4}\,.
\end{align}
Like in Figure~\ref{KV-dispersion-}(dotted line), this gradient model 
leads to a general dispersion with a limitted frequency range for
transmission of waves.

\subsection{Stress gradients}\label{sec-Brenner}

The basic  Hooke  merely elastic rheology described by the non-dispersive hyperbolic
model $\varrho\DDt u={C}u_{xx}$, written
in the rate stress/velocity form as the system $\varrho\Dt v=\sigma_x$ and
$\Dt \sigma={C}v_x$, can be modified by a conservative gradient as
\begin{align}\label{Brenner}
\varrho\Dt{v}=\sigma_x\ \ \ \ \ \text{ and }\ \ \ \ \
\Dt\sigma-\ell^2\Dt\sigma_{xx}={C}v_x\,.
\end{align}
This is an essence of the stress-gradient theory of Eringen \cite{Erin83DENE} and
Aifantis \cite{Aifa92RGLD}; for comparison cf.\ \cite{AskGit10RCSG}.
Substituting the first equation in \eq{Brenner} differentiated in $x$ into
the second one differentiated in $t$ eliminates $v$ and we obtain the
dispersive wave equation
\begin{align}\label{disp-eq-Brenner}
\varrho\big(\DDt\sigma-\ell^2\DDt\sigma_{xx}\big)-{C}\sigma_{xx}=0\,.
\end{align}
The wave equation \eq{wave-eq} is thus enhanced by the micro-inertia,
cf.\ Remark~\ref{rem-microinert} below.

Actually, the same equation as \eq{disp-eq-Brenner} holds in term of
displacement $u$ when substituting that $v=\Dt u$ and integrating
\eq{disp-eq-Brenner} in time. Like in \eq{hyper-Maxwell-1D-dispers}, one
can see the micro-inertia term $\varrho\ell^2\DDt u_{xx}$. This merely
elastic model is fully conservative, i.e.\ there is no energy dissipation
and thus Q-factor is $+\infty$, but anyhow there is a dispersion. Using the
ansatz $\sigma={\rm e}^{{\rm i}(wt+x/\lambda)}$ as in \eq{dispersion-ansatz}
for \eq{disp-eq-Brenner}, we obtain
\begin{align}\label{disp-eq-Brenner+}
\varrho w^2-\frac{\varrho\ell^2}{\lambda^2}w^2-\frac{{C}}{\lambda^2}=0\,.
\end{align}
From this, we obtain the Kramers-Kronig relations
\begin{align}\label{disp-eq-Brenner++}
\varrho(\omega^2-\gamma^2)\Big(1-\frac{\ell^2}{\lambda^2}\Big)
=\frac{{C}}{\lambda^2}\ \ \ \ \ \text{ and }\ \ \ \ \ 2\varrho\gamma
\Big(1-\frac{\ell^2}{\lambda^2}\Big)=0\,.
\end{align}
The latter relation gives $\gamma=0$ so that the former relation results to
\begin{align}
v=\omega\lambda=\sqrt{\frac{{C}}{\varrho(1{-}\ell^2/\lambda^2)}}\,.
\label{v-Brenner}\end{align}
The real-valued velocity is obtained for $\lambda>\lambda_{\rm crit}^{}:=\ell$.
Note that $v$ in \eq{v-Brenner} decays with increasing $\lambda$,
which means the anomalous dispersion. For ultra-long waves, the velocity
asymptotically approaches the velocity in the nondispersive model $\sqrt{{C}/\varrho}$
while for $\lambda\to\lambda_{\rm crit}^{}$, the velocity blows up to $+\infty$.

One can consider a combination of this fully conservative model with the previous
viscous dissipative models. The {\it Kelvin-Voigt rheology with stress
diffusion} would expand \eq{Brenner} as
\begin{align}\label{Brenner-dissip}
\varrho\Dt{v}=\sigma_x\ \ \ \ \ \text{ and }\ \ \ \ \
\big(\Dt\sigma-\ell^2\Dt\sigma_{xx}\big)={C}v_x+\Kv\Dt v_x\,.
\end{align}
This gives the additional term $\Kv\Dt\sigma_{xx}$ on the right-hand side of
\eq{disp-eq-Brenner} and ${\rm i}\Kv w/\lambda^2$ in \eq{disp-eq-Brenner+}.
Therefore \eq{disp-eq-Brenner++} expands as
\begin{align}\label{disp-eq-Brenner+++}
\varrho(\omega^2-\gamma^2)\Big(1-\frac{\ell^2}{\lambda^2}\Big)
=\frac{{C}}{\lambda^2}-\frac{\Kv\gamma}{\lambda^2}\ \ \ \ \ \text{ and }\ \ \ \ \ 2\varrho\gamma
\Big(1-\frac{\ell^2}{\lambda^2}\Big)=\frac{\Kv}{\lambda^2}\,.
\end{align}
The latter relation gives $\gamma$, and then the former relation yields the velocity:
\begin{align}\label{disp-eq-Brenner++++}
v=\omega\lambda=\sqrt{\frac{{C}-\gamma\Kv+\lambda^2\gamma^2}{\varrho(1-\ell^2/\lambda^2)}}
\ \ \ \text{ with }\ \gamma=\frac{\Kv}{2\varrho(\lambda^2{-}\ell^2)}\,.
\end{align}
For $\Kv=0$ we obtain the model \eq{Brenner} with the critical wave-length
$\lambda_\text{\sc crit}^{}=\ell$ while for $\ell=0$ we obtain the Kelvin-Voigt model \eq{dispersion-}
with the critical wave-length $\lambda_\text{\sc crit}^{}=\Kv/(2\!\sqrt{\varrho{C}})$.
Depending which model dominates, i.e.\ whether $\Kv/(2\!\sqrt{\varrho{C}})<\ell$ or not,
the dispersion is monotone (anomalous) or non-monotone, cf.\
Figure~\ref{fig-stress-gradient}-left while the corresponding Q-factor
$1/(1{-}{\rm e}^{-2\lambda\gamma/v})$ is in Figure~\ref{fig-stress-gradient}-right.
\begin{figure}[ht]\hspace*{5em}
\begin{tikzpicture}  
    \begin{axis}[width=6.5 cm,height=5.8 cm,
        xmin=-0.3,xmax=8.5,ymin=-0.1,ymax=3.2,  
        clip=true,  
        axis lines=center,  
        grid = major,  
        ytick={0, 1, 2, 3},  
        xtick={0, 2,...,8},  
        xlabel=$\dfrac1k$, ylabel={$v=v(1/k)$ from \eq{disp-eq-Brenner++++}\hspace*{-7em}},  
        every axis y label/.style={at=(current axis.above origin),anchor=south},  
        every axis x label/.style={at=(current axis.right of origin),anchor=west},  
      ]  
\addplot[very thick,domain=1:8,samples=70]                {sqrt(max(0,(1+4.0^2*(-1/(2*x^2-2*1)+x^2/(2*x^2-2*1)^2))/(1-1^2/x^2)))};
\addplot[very thick,domain=1:8,samples=100,densely dashed]{sqrt(max(0,(1+3.0^2*(-1/(2*x^2-2*1)+x^2/(2*x^2-2*1)^2))/(1-1^2/x^2)))};
\addplot[very thick,domain=1:8,samples=50,dotted]{sqrt(max(0,1/(1-1^2/x^2)))};
 \legend{
      {\scriptsize $\Kv=4$},
      {\scriptsize $\Kv=3$},
      {\scriptsize $\Kv=0$},};
            \end{axis}  
  \end{tikzpicture}\hspace*{.5em}
\begin{tikzpicture}  
    \begin{axis}[width=6.5 cm,height=5.7 cm,
        xmin=-0.3,xmax=8.5,ymin=0.99,ymax=3.2,  
        clip=true,  
        axis lines=center,  
        grid = major,  
        ytick={1, 2, 3},  
        xtick={0, 2,...,8},  
        xlabel=$\dfrac1k$, ylabel={$Q=Q(1/k)=1/(1{-}{\rm e}^{-2\gamma/( kv(1/k)})$\hspace*{-11em}},
        every axis y label/.style={at=(current axis.above origin),anchor=south},  
        every axis x label/.style={at=(current axis.right of origin),anchor=west},  
      ]  
\addplot[very thick,domain=1:8,samples=70]                {1/(1-exp(-2*x*4/((2*x^2-2*1)*sqrt(max(0,(1+4^2*(-1/(2*x^2-2*1)+x^2/(2*x^2-2*1)^2))/(1-1^2/x^2))))))};
\addplot[very thick,domain=1:8,samples=70,densely dashed] {1/(1-exp(-2*x*3/((2*x^2-2*1)*sqrt(max(0,(1+3^2*(-1/(2*x^2-2*1)+x^2/(2*x^2-2*1)^2))/(1-1^2/x^2))))))};
\legend{{\scriptsize $\Kv=4$},
        {\scriptsize $\Kv=3$},
            };
            \end{axis}  
  \end{tikzpicture}\hspace*{-1em}
\caption{
{\sl An anomalous or nonmonotone dispersion of the wave velocity (left) due to 
  \eq{v-Brenner} and \eq{disp-eq-Brenner++++} and the Q-factor (right)
in dependence on the  angular  wavelength $\lambda=1/k$; ${C}=1$, $\varrho=1$, and $\ell=1$, For \eq{v-Brenner} which is
\eq{disp-eq-Brenner++++} with $\Kv=0$, $Q=+\infty$ is not depicted.
}}
\label{fig-stress-gradient}
\end{figure}
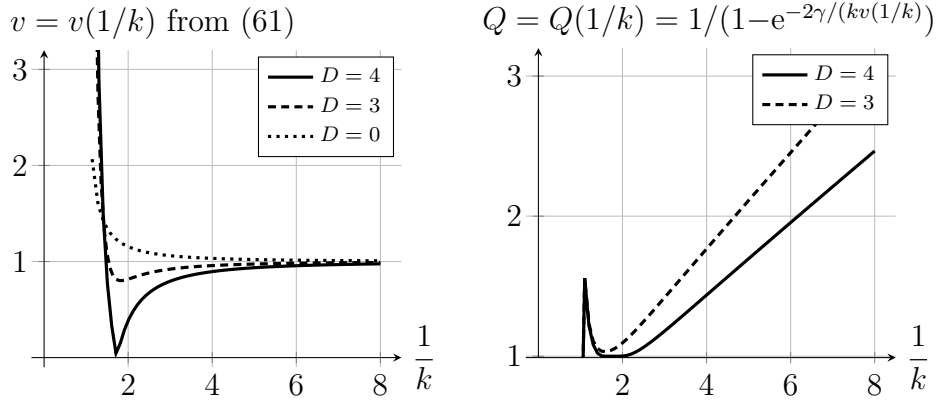

As already mentioned, for $\ell\to0$, we obtain in the limit the conventional
Kelvin-Voigt rheology. The asymptotic behaviour and comparison with that
Kelvin-Voigt model is illustrated in Figure~\ref{fig-stress-gradient+}.

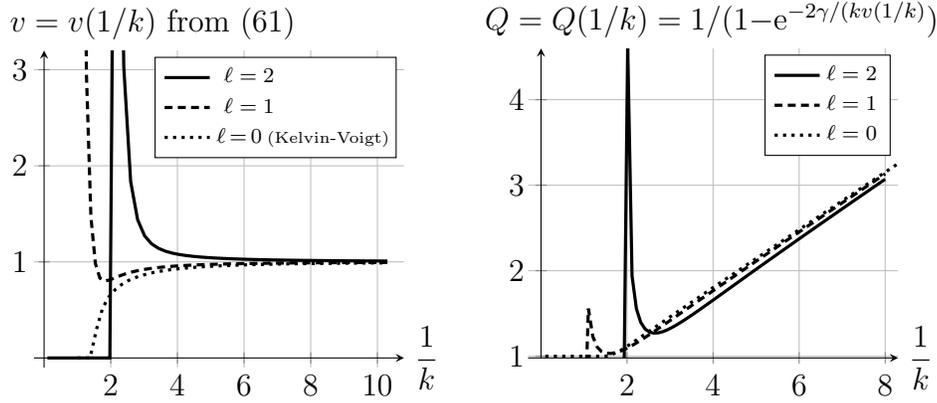
\begin{figure}[ht]\hspace*{5em}
\begin{tikzpicture}  
    \begin{axis}[width=6.5 cm,height=5.8 cm,
        xmin=-0.3,xmax=10.8,ymin=-0.1,ymax=3.2,  
        clip=true,  
        axis lines=center,  
        grid = major,  
        ytick={0, 1, 2, 3},  
        xtick={0, 2,...,8,10},  
        xlabel=$\dfrac1k$, ylabel={$v=v(1/k)$ from \eq{disp-eq-Brenner++++}\hspace*{-7em}},  
        every axis y label/.style={at=(current axis.above origin),anchor=south},  
        every axis x label/.style={at=(current axis.right of origin),anchor=west},  
      ]  
\addplot[very thick,domain=.1:10.3,samples=50]               {sqrt(max(0,(1+3^2*(-1/(2*x^2-2*2.0^2)+x^2/(2*x^2-2*2.0^2)^2))/(1-2.0^2/x^2)))};
\addplot[very thick,domain=1:10.3,samples=70,densely dashed]{sqrt(max(0,(1+3^2*(-1/(2*x^2-2*1.0^2)+x^2/(2*x^2-2*1.0^2)^2))/(1-1.0^2/x^2)))};
\addplot[very thick,domain=1:10.3,samples=50,dotted]        {sqrt(max(0,(1-3^2/(4*x^2))))};
 \legend{
      {\scriptsize $\ell=2$\hspace*{4.5em}},
      {\scriptsize $\ell=1$\hspace*{4.5em}},
     {\scriptsize $\ell\,{=}\,0$\:{\smaller(Kelvin-Voigt)}\hspace*{-.3em}},};
            \end{axis}  
  \end{tikzpicture}\hspace*{.5em}
\begin{tikzpicture}  
    \begin{axis}[width=6.5 cm,height=5.7 cm,
        xmin=-0.3,xmax=8.3,ymin=0.98,ymax=4.6,  
        clip=true,  
        axis lines=center,  
        grid = major,  
        ytick={1, 2, 3, 4},  
        xtick={0, 2,...,8},  
         xlabel=$\dfrac1k$, ylabel={$Q=Q(1/k)=1/(1{-}{\rm e}^{-2\gamma/( kv(1/k)})$\hspace*{-11em}},
        every axis y label/.style={at=(current axis.above origin),anchor=south},  
        every axis x label/.style={at=(current axis.right of origin),anchor=west},  
      ]  
\addplot[very thick,domain=1:8,samples=70]               {1/(1-exp(-2*x*3/((2*x^2-2*2^2)*sqrt(max(0,(1+3^2*(-1/(2*x^2-2*2.^2)+x^2/(2*x^2-2*2^2)^2))/(1-2.0^2/x^2))))))};
\addplot[very thick,domain=1:8,samples=70,densely dashed]{1/(1-exp(-2*x*3/((2*x^2-2*1)*sqrt(max(0,(1+3^2*(-1/(2*x^2-2*1)+x^2/(2*x^2-2*1)^2))/(1-1.^2/x^2))))))};
\addplot[very thick,domain=0.1:14.5,samples=50,dotted]   {1/(1-exp(-3/(x*sqrt(max(0.0001,1-(3/(2*x))^2)))))};
\legend{{\scriptsize $\ell=2$\hspace*{.0em}},
        {\scriptsize $\ell=1$},
        {\scriptsize $\ell=0$},
             };
            \end{axis}  
\end{tikzpicture}\hspace*{-1em}
\caption{
{\sl A normal or nonmonotone dispersion of the wave velocity (left) due to 
  \eq{v-Brenner} and \eq{disp-eq-Brenner++++} and the Q-factor (right)
   in dependence on the  angular  wavelength $\lambda=1/k$; ${C}=1$, $\varrho=1$,
  and $\Kv=3$. For $\ell=0$,
it coincides with the Kelvin-Voigt model from Figure~\ref{KV-dispersion-}.
}}
\label{fig-stress-gradient+}
\end{figure}

The concept of {\it stress diffusion applied to the Maxwell rheology} leads to
the system
\begin{align}\nonumber\\[-2.7em]
\varrho\Dt{v}=\sigma_x\ \ \ \ \ \text{ and }\ \ \ \ \
\frac{\Dt{\sigma}}C+\frac\sigma{D}=v_x+\frac{\ell^2\sigma_{xx}}{D}\,.
\label{Max-stress-diffuse}\end{align}
Realizing the Hooke law $\sigma={C}e$, \eq{Max-stress-diffuse} turns into a system
for $(v,e)$ with a strain diffusion:
\begin{align}
\varrho\Dt{v}={C}e_x\ \ \ \ \ \text{ and }\ \ \ \ \
D\Dt{e}+{C}e=Dv_x+\ell^2{C}e_{xx}\,.
\label{Max-strain-diffuse}\end{align}
Eliminating $v$ from \eq{Max-stress-diffuse}, we obtain the dispersive wave equation
\begin{align}
\frac{\DDt{\sigma}}{C}+\frac{\Dt{\sigma}}{D}-\frac{\sigma_{xx}}{\varrho}
=\frac{\ell^2}{D}\Dt\sigma_{xx}\,.
\end{align}
Using the ansatz \eq{dispersion-ansatz} for $\sigma$, we obtain
$w^2/{C}-{\rm i}w/D-1/(\varrho\lambda^2)={\rm i}\ell^2w/(D\lambda^2)$.
In terms of the real-valued coefficients $\omega$ and $\gamma$, it gives
\begin{align}
\frac{\omega^2{-}\gamma^2}{C}+{\rm i}\frac{2\omega\gamma}{C}
-{\rm i}\frac\omega\Kv+\frac\gamma\Kv-\frac1{\varrho\lambda^2}=
{\rm i}\frac{\ell^2}{D}\frac{\omega}{\lambda^2}
-\frac{\ell^2}{D}\frac{\gamma}{\lambda^2}\,,
\end{align}
The Kramers-Kronig relations \eq{KK-relation+} expand to
\begin{align}\label{KK-relation+diffuse}
&\frac1{C}(\omega^2-\gamma^2)+\frac1D\gamma-
\frac1{\varrho\lambda^2}+\frac{\ell^2}{D}\frac{\gamma}{\lambda^2}=0
\ \ \ \ \text{ and }\ \ \ \
\frac2{C}\gamma=\frac1D+\frac{\ell^2}{D\lambda^2}\,,
\end{align}
After some algebra, we obtain the velocity
\begin{align}\label{v-stress-diffusion-Max}
v(\lambda)=\omega^2\lambda^2=\sqrt{\,\frac {C}{\varrho}-
\frac{{C}^2(\lambda^2{+}\ell^2)^2}{4\Kv^2\lambda^2}}\,.
\end{align}
This yields a general nonmonotone dispersion which is, together with the
corresponding Q-factor $1/(1-{\rm e}^{-2\lambda\gamma/v})$ as in
\eq{def-Q-factor}, displayed in Fig.\,\ref{fig-stress-diff-Max}.
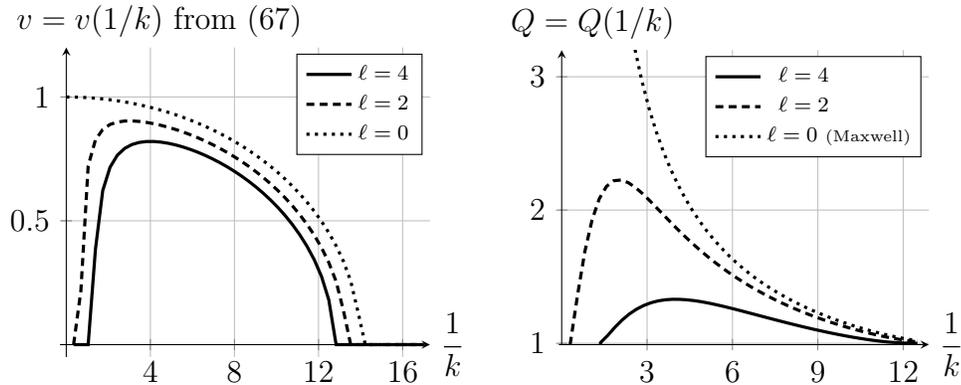
\begin{figure}[ht]\hspace*{5em}
\begin{tikzpicture}  
    \begin{axis}[width=6.5 cm,height=5.7 cm,
        xmin=-0.3,xmax=17.3,ymin=-0.05,ymax=1.2,  
        clip=true,  
        axis lines=center,  
        grid = major,  
        ytick={0, 0.5, 1},  
        xtick={0, 4,...,20},  
        xlabel=$\dfrac1k$, ylabel={$v=v(1/k)$ from \eq{v-stress-diffusion-Max}\hspace*{-6em}},
    every axis y label/.style={at=(current axis.above origin),anchor=south},  
    every axis x label/.style={at=(current axis.right of origin),anchor=west},  
      ]  
\addplot[very thick,domain=0:17,samples=50]{sqrt(max(0,1-1*(x^2+4^2)^2/(4*x^2*7^2)))};
\addplot[very thick,domain=0:17,samples=50,densely dashed]{sqrt(max(0,1-1*(x^2+3^2)^2/(4*x^2*7^2)))};
\addplot[very thick,domain=0:17,samples=50,dotted]{sqrt(max(0,1-1*x^2/(4*7^2)))};
 \legend{{\scriptsize $\ell=4$\hspace*{0em}},
      {\scriptsize $\ell=2$\hspace*{0em}} ,
      {\scriptsize $\ell=0$\hspace*{0em}} ,
             };
            \end{axis}  
  \end{tikzpicture}\hspace*{.5em}
\begin{tikzpicture}  
    \begin{axis}[width=6.5 cm,height=5.5 cm,
      xmin=-0.1,xmax=12.9,ymin=.99,ymax=3.2,  
        clip=true,  
        axis lines=center,  
        grid = major,  
        ytick={1, 2, 3},  
        xtick={0, 3,...,12},  
         xlabel=$\dfrac1k$, ylabel={$Q=Q(1/k)$\hspace*{-2em}},
   every axis y label/.style={at=(current axis.above origin),anchor=south},  
   every axis x label/.style={at=(current axis.right of origin),anchor=west},  
      ]  
 \addplot[very thick,domain=.1:12.5,samples=50]              {1/(1-exp(-x*(x^2+4^2)/(7*x^2*sqrt(max(0,1-1*(x^2+4^2)^2/(4*x^2*7^2))))))};
\addplot[very thick,domain=.05:12.5,samples=50,densely dashed]{1/(1-exp(-x*(x^2+2^2)/(7*x^2*sqrt(max(0,1-1*(x^2+2^2)^2/(4*x^2*7^2))))))};
\addplot[very thick,domain=0:12.5,samples=50,dotted]{1/(1-exp(-1*x/(7*sqrt(max(0.0001,1-1*x^2/(4*7^2))))))};
\legend{{\scriptsize $\ell=4$\hspace*{3em}},
      {\scriptsize $\ell=2$\hspace*{3em}} ,
  {\scriptsize $\ell=0$  \tiny(Maxwell)\!} ,
            };
            \end{axis}  
\end{tikzpicture}\hspace*{-1em}
\caption{
{\sl A nonmonotone dispersion of the wave velocity (left) due to 
  \eq{v-stress-diffusion-Max} and the Q-factor (right)
   in dependence on the  angular  wavelength $\lambda=1/k$;
  ${C}=1$, $\varrho=1$, and $\Kv=7$. For $\ell=0$,
  it coincides with the Maxwell model from Figure~\ref{KV-dispersion-}.
}}
\label{fig-stress-diff-Max}\end{figure}

\subsection{Gradient regularization of the kinematic constraint}
\label{sec-kinematic-grad}

Another possibility is to involve the dissipative stress gradient into the
kinematic constraint $\Dt e=v_x$ in \eq{wave-eq}, leading to the
 parabolic-type regularization
\begin{align}\label{diffusive-regularization}
\Dt e=v_x+\varepsilon e_{xx}
\end{align}
with $\varepsilon>0$ a (pressumably small) diffusion parameter (in m$^2$/s). 
 Alternatively, one can think about a similar regularization
by a stress gradient instead of the strain gradient, i.e.
\begin{align}\label{diffusive-regularization+}
\Dt e=v_x+\varepsilon\sigma_{xx}
\end{align}
with some (presumably small) coefficient $\varepsilon>0$ in
m$^2$/(Pa\,s)=m$^5$/(J\,s). Such a dissipative modification
of the purely conservative model \eq{wave-eq} reads as
\begin{align}\label{Brenner++}
\varrho\Dt{v}=\sigma_x\ \ \ \text{ with }\ \ \ \sigma={C}e
\ \ \ \ \ \text{ and }\ \ \ \ \
\Dt{e}=v_x+\varepsilon\sigma_{xx}\,.
\end{align}
The energetics behind \eq{Brenner++} can be revealed (assuming homogeneous
boundary conditions as in Remark~\ref{rem-KV-hyper-enrg}) by testing $\varrho\Dt v={C}e_x$ by $v$.
When integrating it over a spatial domain $\varOmega$, this gives \
\begin{align*}\frac{\d}{\d t}\int_\varOmega\frac\varrho2 v^2\d x+\int_{\varOmega}\!{C}ev_x\d x
  &=\frac{\d}{\d t}\int_\varOmega\frac\varrho2 v^2\d x
  +\int_\varOmega\!{C}e\big(\Dt{e}-\frac{\sigma_{xx}}\Kv
  \varepsilon\sigma_{xx}\big)\,\d x
\\&=\frac{\d}{\d t}\int_\varOmega\Big(\frac\varrho2 v^2+\frac12{C}e^2\Big)\,\d x
    +\int_\varOmega\varepsilon{C}\sigma_x^2\,\d x=0\,.
\end{align*}
The last term reveals the dissipation rate 
$\varepsilon{C}\sigma_x^2$, which  shows the dissipative
character of this regularization and  explains why
the last term in \eq{Brenner++} is  somtimes  referred as the
stress-diffusion.
Differentiating the first equation in \eq{Brenner++} in space  
and the second one in time allows for the elimination of $v$ to obtain
the dispersive wave equation $\DDt \sigma/{C}-\sigma_{xx}/\varrho
-\varepsilon\Dt \sigma_{xx}=0$. In terms of $e$, it can be written as
\begin{align}\label{Brenner++disp+}
  \varrho\DDt{e}-\varepsilon\varrho{C}\Dt e_{xx}-{C}e_{xx}=0\,,
\end{align}
which has the same structure (and thus the same dispersion and Q-factor)
as in the Kelvin-Voigt model \eq{dispersion-} except that, instead of $\Kv$ in
\eq{dispersion-}, the viscosity coefficient in \eq{Brenner++disp+} is now
$\varepsilon\varrho{C}$.

Such parabolic-type regularization 
can be combined with the previous viscous dissipative
models. In the case of the Kelvin-Voigt model which expands the momentum
equation by viscosity, it leads to 
\begin{align}\label{Brenner-disip}
\varrho\Dt{v}=\sigma_x+\Kv v_{xx}\ \ \ \text{ with }\ \ \ \sigma={C}e
\ \ \ \text{ and }\ \ \ \Dt{e}=v_x+\varepsilon\sigma_{xx}\,.
\end{align}
Differentiation the first equation in \eq{Brenner-disip}
in space and the latter equation both once in time
and separately also twice in space allows for the elimination of $v$ 
to obtain the dispersive wave equation in terms of the stress $\sigma$:
\begin{align}\label{Brenner-disip+}
\varrho\DDt{\sigma}-{C}\sigma_{xx}-\big(\Kv{+}
\varepsilon\varrho{C}\big)\Dt\sigma_{xx}
+\varepsilon{C}\Kv\sigma_{xxxx}=0\,.
\end{align}
It is to observe that, qualitatively, this model exhibits the same dispersion
and Q-factor as the conservative-gradient enhancement \eq{dispersion+}.
In particular, a nondispersive situation appears also here if
$2\varrho{C}\Kv=\varepsilon(\Kv^2{+}\varrho^2{C}^2)$.

An alternative to \eq{Brenner-disip} arises by  using the
total stress $\sigma={C}e+\Kv\Dt{e}$ instead of only the elastic stress  
$\sigma={C}e$. This yields
\begin{align}\label{Brenner-disip-alt}
\varrho\Dt{v}=\sigma_x\ \ \ \text{ with }\ \ \ \sigma={C}e+\Kv\Dt{e}
\ \ \ \text{ and }\ \ \ \Dt{e}=v_x+\varepsilon\sigma_{xx}\,.
\end{align}
Differentiation the first equation in \eq{Brenner-disip-alt}
in space and the latter equation in time allows for the elimination of $v$
to obtain the dispersive wave equation in terms of the strain $e$:
\begin{align}\label{Brenner-disip++}
  \varrho\DDt{e}-{C}e_{xx}-\varepsilon\varrho\Kv\DDt e_{xx}
-\big(\Kv{+}\varepsilon\varrho{C}\big)\Dt e_{xx}=0\,.
\end{align}
Again we can see a micro-inertia term $\DDt e_{xx}$ and, in fact, the
same structure (and thus the  similar  dispersion and Q-factor)
as the Kelvin-Voigt model with  the  stress diffusion \eq{Brenner-dissip}.
Naturally, for $\varepsilon\to0$, both models \eq{Brenner-disip} and
\eq{Brenner-disip-alt} become equivalent to (each other and to) the
Kelvin-Voigt model \eq{dispersion-}.

\begin{remark}[{\sl Micro-inertia}.]\label{rem-microinert}\upshape
The inertia-gradient terms of the type $\varrho\DDt u_{xx}$ as in
\eq{hyper-Maxwell-1D-dispers} or, when  written in terms of $u$,
in \eq{disp-eq-Brenner} or \eq{Brenner-disip++} is sometimes called
micro-inertia.
See e.g.\ \cite{AskAif09GEFW,AskAif11GESD,Jira04NTCM,MNAB17RMIE,MetAsk02ODCG} for
discussion about micro-inertia as invented by R.D.\,Mindlin \cite{Mind64MSLE}
and related possible other gradient terms and dispersion analysis.
Cf.\ also \cite{EngPas03WMSN} or \cite[Sect.6.3-4]{BerVan17IVT}.
The inviscid variant of \eq{Brenner-disip++}, i.e.\ the hyperbolic equation
of the type $\varrho(\DDt u{-}\ell^2\DDt u_{xx})-{C}u_{xx}=0$
as \eq{disp-eq-Brenner}, is sometimes called
a Love-Rayleigh model, cf.\ \cite[Sect.\,1.2.4]{BerVan17IVT}.
\end{remark}

\def\XX{\mbox{\bf{x}}}
\def\upyy{\mbox{\bf{y}}}
\def\upFF{\mbox{\bf{F}}}
\def\smallupFF{\mbox{\footnotesize\bf{F}}}
\def\upSS{\mbox{\bf{S}}}
\def\upDD{\mbox{\bf{D}}}
\def\upCC{\mbox{\bf{C}}}
\def\upRR{\mbox{\bf{R}}}
\def\upvv{\mbox{\bf{v}}}
\def\upOmega{\Omega}

\section{Notes about gradient theories at large strains}\label{sec4}

Models in the large-strain continuum mechanics in the three space
dimensions are inevitably
governed by very nonlinear partial differential equations.
In dynamical situations, rigorous analytical justification of
such models as far as a mere existence of (suitably defined notions
of weak) solutions is problematic, cf.\ \cite{Ball02SOPE,Ball10PPNE}.
There is a certain agreement that involving some gradient theories is
inevitable to facilitate a rigorous analysis and to ensure stability
and convergence of various numerical approximation strategies. This section
aims to survey various options from the perspective of
the dispersive models in the previous section~\ref{sec3}.

In the large-strain continuum mechanics, the basic geometrical concept is a 
{\it deformation} $\upyy:\upOmega\to\R^3$ as a mapping from a
reference configuration $\upOmega\subset\R^3$ into the physical space $\R^3$.
We will denote the referential (resp.\ the  current) quantities by uprighted
(resp.\ slanted) fonts. In particular, we denote by $\XX$ and $\xx$
the reference (Lagrangian) and the  current  (Eulerian) point
coordinates, respectively. The further basic geometrical object is the
(referential) {\it deformation gradient} $\upFF(\XX)=\nabla_{\XX}^{}\upyy$.

If evolving in time, $\xx=\upyy(t,\XX)$ is sometimes called a ``motion''.
The inverse motion $\bm\upxi=\upyy^{-1}:\upyy(\upOmega)\to\upOmega$, if it
exists, is called a {\it return } (or sometimes a {\it reference}) {\it mapping}.
The important quantity is the (referential) velocity
$\upvv=\frac{\d}{\d t}\upyy(t,\XX)$
with $\d/\d t$ the derivative with respect to time of a time dependent
function. When composed with the return mapping $\bm\upxi$, we obtain
the Eulerian representations 
\begin{align}\label{F-v-Eulerian}
\FF(t,\xx)=\upFF(t,\bm\upxi(\xx))\ \ \ \ \text{ and }\ \ \ \
\vv(t,\xx)=\upvv(t,\bm\upxi(\xx))\,.
\end{align}

The rheologies from Section~\ref{sec2}, exploiting one elastic element,
are now characterized by a non-quadratic stored energy
$\varphi=\varphi(F)$ with $F\in\R^{3\times3}$, being assumed frame indifferent,
i.e.\ a function of the right Cauchy-Green tensor $C=F^\top F$ and
hence surely nonconvex. Moreover, the blow-up $\varphi(F)\to+\infty$
when $\det F\to0+$ is often imposed to grant a local non-interpenetration
of the deformed medium. Existence of such potential of conservative
forces is an essence of the concept of {\it hyperelastic materials}.
Among several options, we specifically consider $\varphi$ in J/m$^3$=Pa
(not in  J/kg), meaning energy per referential (not  current) volume.

\subsection{Lagrangian formulation}

The small-strain  1-dimensional  kinematic constraint $\Dt e=v_x$ in \eq{wave-eq}
is now ``translated'' into the large-strain Lagrangian formulation as
\begin{align}\label{kinem-Lagrange}
\Dt\upFF=\nabla\upvv\,.
\end{align}
The elastic response now leads to the first {\it Piola-Kirchhoff stress}
$\upSS=\varphi'(\upFF)$,
so that the large-strain analog of the fully conservative model \eq{wave-eq}
is now $\rho\Dt \upvv-{\rm div}\upSS=\bm0$ with $\upSS=\varphi'(\upFF)$
and with a referential mass density $\rho=\rho(\XX)$
to be completed by \eq{kinem-Lagrange}.

The large-strain Kelvin-Voigt model \eq{dispersion+} uses also
a nonlinear dashpot governed by a dissipative potential $\zeta=\zeta(\upFF,\Dt \upFF)$.
The frame indifference of dissipative forces is quite delicate, as pointed out in
\cite{Antm98PUVS}, and dictates that $\zeta$ should be a function of
$\upCC=\upFF^\top\!\upFF$ and its rate
$\Dt \upCC=2{\rm sym}(\upFF^\top\!\Dt \upFF)$. The analog to the  1-dimensional 
linear dashpot $\sigma=\Kv\Dt e$ is now $4\Kv \upFF{\rm sym}(\upFF^\top\!\Dt \upFF)$. 
In simple materials, as pointed out also in \cite[Sec.\,3.2]{Ball02SOPE},
this model does not seem mathematically tractable as far as existence of
global solutions or their approximation. Actually,
some results are available only for
short-time analysis or usage of a certain very weak (measure-valued) concept of solutions
or on a non-invariant viscous stress; cf.\
\cite{Demo00WSCN,DeStTz01VAST,Proh08CFEB,Rieg03YMSN,Tved08QEVR}.

The conservative-gradient capillarity-type enhancement of the Kelvin-Voigt model
like \eq{dispersion+} consists in involving the conservative hyperstress
$\ell^2{C}\nabla\upFF$, i.e.\ the conservative stress
$-{\rm div}(\ell^2{C}\nabla\upFF)$.
This can be obtained from the extended stored energy
$\varphi(\FF)+\frac12\ell^2{C}{\mid}\nabla\FF{\mid}^2$.
The large-strain analog of the model \eq{dispersion+} now reads as 
\begin{subequations}\label{KV-Lagrange}
\begin{align}\label{KV-Lagrange1}
\!\rho\Dt\upvv-{\rm div}\big(\upDD{+}\upSS\big)=\bm0
\ &\text{ with }\ \upSS=\varphi'(\upFF){-}{\rm div}(\ell^2{C}\nabla\upFF) 
\ \\[-.3em]&\text{ and }\ \ \upDD=4\Kv\upFF{\rm sym}\Big(\upFF^\top\Dt{\upFF}\Big)\!
\label{KV-Lagrange2}\end{align}\end{subequations}
to be combined with the kinematic equation \eq{kinem-Lagrange}.
The conservative variant, i.e.\ \eq{KV-Lagrange} with $\Kv=0$,
gives the purely elastic model analogous to 
\eq{dispersion+} with $\Kv=0$ as in Figure~\ref{Hook-dispersion-anomalous},
was analytically justified in \cite[Sect.9.2.1]{KruRou19MMCM}.
The Kelvin-Voigt model \eq{dispersion+} with $\Kv>0$ was analyzed in
\cite{MieRou20TKVR} or \cite[Sect.9.3]{KruRou19MMCM}.
Conceptually, according to the analysis from Section~\ref{sec-KV-cons},
this model \eq{KV-Lagrange} has a potentiality to suppress the dispersion of
elastic waves.

The analog of the small-strain Green-Naghdi additive decomposition
\eq{Maxewell-internal-par} is now
conventionally replaced by the Kr\"oner-Lee-Liu \cite{Kron60AKVE,LeeLiu67FSEP}
{\it multi\-pli\-cative decomposition} $\upFF=\upFF\EL\upFF\IN$ with
$\upFF\EL$ the elastic  distortion $\upFF\IN$  and the inelastic
distortion $\upFF\IN$  which is considered isochoric, i.e.\
$\det\upFF\IN=1$; this constraint allows to distinguish the deviatoric
and the isochoric (volumetric) parts of the model and to
considered different rheologies for them, typically a solid
rheology (often even ideally rigid, so incompressible) for the
volumetric part while some fluidic rheology for the deviatoric part.
The stored energy $\varphi$ now depends on
$\upFF\EL=\upFF\upFF\IN^{-1}$ instead of $\upFF$. 
In the case of Maxwell's rheology with a linear creep,
the other ingredient is the 
dissipation potential $\zeta=\zeta(\upFF\IN,\cdot):\R^{3\times 3}\to\R$
depending (in the case of linear creep quadratically) on the inelastic distortion
rate $\Dt\upFF\IN$.

The large-strain analog of the Jeffreys rheology from Section~\ref{sec2.3} now 
leads to the system of 1st-order differential equations for $(\upvv,\upFF,\upFF\IN)$:
\begin{subequations}\label{Max-Lagrange}
\begin{align}\nonumber
    &\!\rho\Dt\upvv-{\rm div}\big(\upDD{+}\upSS\upFF\IN^{-\top}\big)
  =\bm0\ \ \ \text{ with }\ \,\upSS=\varphi'(\upFF\EL)
    \\[-.3em]&\hspace{11.5em}\label{Max-Lagrange1}\ \text{ and }\ \ 
    \upDD=4\Kv_1\upFF{\rm sym}\big(\upFF^\top\Dt{\upFF}\big)\,,
\\[-.2em]&\!
\zeta_{\Dt\smallupFF\IN}'\!\!\big(\upFF\IN;\!\!\Dt\upFF\IN\big)
+\upSS\IN=\bm{0}\ \ \ \ \text{ with }\:\upSS\IN\!\!
=\upFF^\top\!\upSS(\upFF\IN^{-1})'\:\text{ with }\:\upFF\EL\!\!
=\upFF\upFF\IN^{-1}\!\!
\label{Max-Lagrange2}\end{align}\end{subequations}
to be combined with the kinematic equation \eq{kinem-Lagrange}. 
In \eq{Max-Lagrange2}, the differential of the nonlinear mapping
$(\cdot)^{-1}=:\R^{3\times3}\to\R^{3\times3}$ is the 4th-order tensor
$(\upFF\IN^{-1})'={\rm Cof}'\upFF\IN^\top-{\rm Cof}\upFF\IN^\top{\otimes}{\rm Cof}\upFF\IN$
with ${\rm Cof}'$ denoting the differential of the cofactor-mapping
${\rm Cof}=(\det\cdot)(\cdot)^{-\top}:\R^{3\times3}\to\R^{3\times3}$; here $\det\upFF\IN=1$ is used. 
For $\Kv_1=0$, this degenerates to a nonlinear hyperbolic system
(representing the analog of the Maxwell rheology from Section~\ref{sec2.2})
and its solution is analytically problematic. Even, the
``parabolic'' situation $\Kv_1>0$ does not seem much better at this point. 

The conservative gradient en\-han\-cement like \eq{hyper-Maxwell-1D-conserve}
would be now ``translated'' into large strains 
by expanding $\upSS$ in \eq{Max-Lagrange1} as in \eq{KV-Lagrange1}, i.e.\ here 
$\upSS=\varphi'(\upFF\EL){-}{\rm div}(\ell_1^2{C}\nabla\upFF\EL)$.
This corresponds to the enhancement of the stored energy $\varphi(\upFF\EL)$
by the term $\frac12\ell_1^2{C}{\mid}\nabla\upFF\EL{\mid}^2$.
Yet, the rigorous analysis seems, however, again analytically open,
cf.\ \cite[Remark~9.4.6]{KruRou19MMCM} or \cite[Remark~2.3]{RouSte19FTCS}.
Here a combination with the dissipative gradient like in Sections~\ref{sec-Max-diss}
but of a higher order can help: 
with $\zeta$ in \eq{Max-Lagrange2} expanded by the
2nd-order gradient $\frac12\int_\Omega{\mid}\ell^2\nabla^2\Dt \upFF\IN{\mid}^2\,\d x$
which, when the derivative of $\zeta$ in \eq{Max-Lagrange2} understood in
a functional sense, contributes to the left-hand side of \eq{Max-Lagrange2} by the 
4th-order-gradient stress ${\rm div}^2(\ell^4\nabla^2\Dt \upFF\IN)$. For a rigorous
analysis, we refer to \cite{DaRoSt21NHVM}.

One can also consider a conservative gradient as in the
Poynting-Thomson standard solid in Remark~\ref{rem-Zener}, here
enhancing $\upSS\IN$ in \eq{Max-Lagrange2} by the term
$-{\rm div}(H\nabla\FF\IN)$. In \cite{RouSte18TEPR,RouSte19FTCS} and
in \cite[Sect.9.4]{KruRou19MMCM}, a simplified enhancement
by $\frac12\ell^2{C}{\mid}\nabla\upFF{\mid}^2$, leading to the
higher-order gradient contribution to the Piola-Kirchhoff stress
$-\upFF{\rm div}(\ell^2{C}\nabla\upFF)$, was rigorously analyzed 
but, mixing $\upFF\EL$ and $\upFF$ in the stored energy is rather
a misconception for a Maxwellian-type rheologies, corresponding rather
to Zener's standard solid model. 

\begin{remark}[{\sl Normal dispersion by multipolar viscosity}.]\label{rem-normal}\upshape
The multipolar viscosity $\ell^2\Kv\Dt u_{xxxx}^{}$
in Sections~\ref{sec-KV-diss} or  Sect.~\ref{sec-KV-mixed}
should now involve the gradient of the rate of $\upFF$ in the 
the dissipation potential. Yet, by frame indifference the dissipation potential
should depend on the rate $\upRR$ of the right Cauchy-Green tensor
$\upCC=\upFF^\top\upFF$, i.e.\ on $\upRR=\Dt \upCC=2{\rm sym}(\upFF^\top\!\Dt \upFF)$,
rather than on $\Dt \upFF$. For the quadratic isotropic dissipation potential
$\mathscr{D}(\upRR)=\frac12\ell^2\Kv\int_\Omega{\mid}\nabla\upRR{\mid}^2\,\d\XX$,
the contributions to the dissipative Piola-Kirchhoff stress $\upDD$ arising
as the functional derivative of $\Dt \upFF\mapsto\mathscr{D}(\upRR)$ when realizing that 
\begin{align*}
\nabla\upRR=\nabla\Dt{\overline{(\upFF^\top\upFF)}}
=\nabla\upFF^\top\!{\cdot}\Dt\upFF+\upFF^\top\!{\cdot}\nabla\Dt\upFF+
\nabla\Dt{\upFF}^\top{\cdot}\upFF
+\Dt{\upFF}^\top{\cdot}\nabla\upFF\,
\end{align*}
consists both from the stress depending linearly on $\Dt \upFF$ through
coefficients of the type $\nabla\upFF{\otimes}\nabla\upFF$ and the divergence
of a hyperstress depending linearly on $\nabla\Dt \upFF$ through 
coefficients of the type $\upFF{\otimes}\upFF$. The rigorous analysis
(as well as convergent numerical strategies) of such model
seems very nontrivial because it needs strong convergence
(of an approximation) of $\nabla\upFF$, and open.
\end{remark}

\subsection{Eulerian formulation}

The Eulerian velocity $\vv$ from \eq{F-v-Eulerian} is employed in the
{\it convective time derivative} $\DT{}(\bm\cdot)
=\frac{\pl}{\pl t}(\bm\cdot)+(\vv{\cdot}\nabla)(\bm\cdot)$ with $\nabla$
considered  in this subsection  with respect to the  current 
coordinates, to be used for scalars and, component-wise, for vectors or tensors.
The small-strain kinematic constraint $\Dt e=v_x$ in \eq{wave-eq} is now ``translated'' into
large-strain Eulerian formulation as
\begin{align}
\DT\FF=(\nabla\vv)\FF\,.
\label{kinem-Euler}\end{align}
Equivalently, in terms of the {\it left Cauchy-Green tensor} $\BB=\FF\FF^\top\!$,
\eq{kinem-Euler} can be written as
\begin{align}
\DT\BB=(\nabla\vv)\BB+\BB(\nabla\vv)
\label{kinem-Euler+}\end{align}
or, written shortly, $\OLD\BB=\bm0$ where $\OLD\BB:=\DT{}\BB-(\nabla\vv)\BB-\BB(\nabla\vv)$
denotes the so-called {\it upper-convected time derivative}.

In contrast to the referential mass density $\rho=\rho(\XX)$,  
the  current  mass density $\varrho=\varrho(t,\xx)$ now depends also on time
and its evolution is governed by the continuity equation
\begin{align}\nonumber\\[-2.7em]\label{cont-eq}
  \DT\varrho=({\rm div}\,\vv)\varrho\,.
\end{align}
When the initial deformation is identity, i.e.\ $\xx=\upyy(0,\XX)=\XX$ and
\eq{cont-eq} is completed with the initial condition $\varrho(0,\xx)=\rho(\XX)$,
it holds $\varrho=\rho/\!\det\FF$.
The (hyper)elastic response is again governed by the stored energy $\varphi$
which is now a function of the Eulerian deformation gradient $\FF$. This
leads to the symmetric {\it Cauchy stress} $\bm{T}=\varphi'(\FF)\FF^\top\!/\!\det\FF$.
The momentum equation is now $\varrho\DT{}\vv={\rm div}\,\bm{T}$ or, equivalently,
$\Dt\vv={\rm div}(\bm{T}{-}\varrho\vv{\otimes}\vv)$.
Analysis of this nonlinear hyperbolic system is naturally very difficult and 
only limited results are available in special cases, cf.\ e.g.\
\cite{HuWan12FSCV,SidTho05GETD}.

The analog of the 1D small-strain Kelvin-Voigt model \eq{dispersion-} written as
$\varrho\Dt v-(\Kv v_x{+}{C}e)_x^{}=0$ now leads to the system for $(\varrho,\FF,\vv)$
composed from \eq{kinem-Euler}, \eq{cont-eq}, and the momentum equation
\begin{subequations}\label{KV-Euler}\begin{align}\label{KV-Euler1}
  \varrho\DT\vv-{\rm div}\big(\bm{T}{+}\bm{D}\big)=\bm0\ \ &\text{ with }\
  \bm{T}=\frac{\varphi'(\FF)\FF^\top\!\!}{\det\FF}\
  \\&\text{ and }\;\ \bm{D}=\Kv\ee(\vv)\,.
  \label{KV-Euler2}\end{align}\end{subequations}
The mere existence of suitably defined solutions (not speaking about
 its uniqueness or  numerical approximation) of this (very nonlinear) system is
 nowadays generally considered as still open if no gradient theories are
involved. Some results exist only for 
vanishing shear elastic response (i.e.\ for viscoelastic fluids
rather than solids); we refer to \cite{Feir04DVCS,FeNoPe01EGDW,HuMas17GSRH}.

The dissipative-gradient extension like \eq{dispersion} here uses
\eq{KV-Euler1} with the dissipative stress
$\bm{D}=\Kv\ee(\vv)-{\rm div}(\ell^2\Kv\nabla\ee(\vv))$ or
with $\bm{D}=\Kv\ee(\vv)-{\rm div}(\ell^2\Kv\nabla^2\vv)$. This 
is the concept of so-called {\it multipolar materials} by
Ne\v cas at al.\ \cite{Neca94TMF,NeNoSi91GSCI,NecRuz92GSIV}
or of solids \cite{Ruzi92MPTM,Silh92MVMS}, later also by Fried and
Gurtin \cite{FriGur06TBBC}, inspired by Toupin \cite{Toup62EMCS} and
Mindlin \cite{Mind64MSLE}, and allowing for rigorous analysis, also used
in \cite{RouSte23VESS}. Conceptually, according to the analysis from
Section~\ref{sec-KV-diss}, the hyper-viscosity in the model \eq{KV-Euler}
has a potentiality to cause lesser dispersion and attenuation than the
usual viscosity for wavelengths bigger than the critical, cf.\
Figure~\ref{KV-dispersion-comparison}.
In a nonlinear variant of the multipolar viscosity, it may grant
boundedness of velocity gradient, which further opens the way to analysis
of the whole system \eq{KV-Euler}, cf.\ \cite{Roub22VELS}.
It is well recognized that without the mentioned 
velocity-gradient boundedness, the singularities
of transported variables $\varrho$ and $\FF$, whose
occurrence in solids may be debatable,
may develop, cf.\ \cite{AlCrMa19LRCE}.

The diffusive regularization of the kinematic constraint \eq{kinem-Euler}
like \eq{diffusive-regularization} now leads to 
\begin{align}
\DT\FF=(\nabla\vv)\FF+\varepsilon\Delta\FF\,.
\label{kinem-Euler-reg}\end{align}
This diffusive regularization \eq{kinem-Euler-reg} used for the Kelvin-Voigt
model \eq{KV-Euler} facilitates the rigorous analysis, as shown in the
incompressible case in \cite{ACGR22CHMC,BFLS18EWSE,GKMS22SWPS}.
Although violation of the ultimate kinematic constraint \eq{kinem-Euler} is
surely not physically legitimate, the linear small-strain analysis based on
the dispersive wave equation \eq{Brenner-disip+} yields a certain modelling
justification to facilitate the propagation of high-frequency waves.
 Similar diffusive regularization could be used for \eq{kinem-Euler+}, leading
to $\DT{}\BB=(\nabla\vv)\BB+\BB(\nabla\vv)+\varepsilon\Delta\BB$.

The Maxwell or the Jeffreys rheologies again use the  Kr\"oner-Lee-Liu 
multi\-pli\-cative decomposition,
now for the Eulerian deformation gradient $\FF=\FF\EL\FF\IN$  again
with the inelastic distortion $\FF\IN$ considered isochoric, i.e.\ $\det\FF\IN=1$.
In the rate form $\nabla\vv=\DT{}(\FF\EL\FF\IN)
=(\DT{}\FF\EL)\FF\IN+\FF\EL\DT{}\FF\IN$
with introducing the inelastic (creep) distortion rate $\LL\IN=(\DT{}\FF\IN)\FF\IN^{-1}$,
we obtain the extension of \eq{kinem-Euler}, devised by Lee \cite{Lee69EPDF}, as
\begin{align}
\DT{\FF\EL}=(\nabla\vv)\FF\EL-\FF\EL\LL\IN\,.
\label{Lee-formula}\end{align}
The dissipative-gradient small-strain Jeffreys model \eq{Jeffreys-small-strain+++grad}
written in the rate formulation by substituting $v=\Dt u$,
i.e.\ using both the gradient term
${\rm div}^2(\ell_1^2\Kv_1\nabla^2\vv)$ and the gradient term
${\rm div}(\ell_2^2\Kv_2\nabla\LL\IN)$ now reads as
\begin{subequations}\label{Max-Euler}\begin{align}\nonumber
&\varrho\DT\vv-{\rm div}\big(\bm{T}{+}\bm{D}\big)=\bm0\ \ \text{ with }\
  \bm{T}=\varphi'(\FF\EL)\FF\EL^\top/\!\det\FF\EL
  \\[-.3em]&\hspace{10.7em}\text{ and }\;\ \bm{D}=\Kv_1\ee(\vv)
  -{\rm div}(\ell_1^2\Kv_1\nabla^2\vv)\,,
  \\[-.3em]&\DT{\FF\EL}=(\nabla\vv)\FF\EL-\FF\EL\LL\IN\,,
  \\[-.3em]&\Kv_2\LL\IN=
  \FF\EL^\top\!\varphi'(\FF\EL)/\!\det\FF\EL+{\rm div}(\ell_2^2\Kv_2\LL\IN)
\label{Max-Euler-c}\end{align}\end{subequations}
together with \eq{cont-eq}. 
For the rigorous analysis of \eq{Max-Euler} but with nonlinear higher-gradient
viscosities, we refer to \cite{Roub22QHLS}. For $\Kv_1=0$,
we obtain the dissipative-gradient enhancement of the Maxwell model from
\eq{sec-Max-diss} but the rigorous analysis of this nonlinear hyperbolic
model seems open.

 The analog of the stress diffusion \eq{Max-stress-diffuse}
is $\DT{}\TT+(C/D)\TT=(\nabla\vv)\TT+\TT(\nabla\vv)+\varepsilon\Delta\TT$.
Combined with the Stokes rheology into a Jeffreys model of the type, it was
analyzed in the 2-dimensional situation in \cite{BaLuSu17ELDFE} and in the 3-dimensional
situation with the nonlinear diffusion and Stokes terms in \cite{KrPoSa15GERM}.
The analog of the Maxwell model with the strain diffusion
\eq{Max-strain-diffuse} in terms of left-Cauchy-Green strain tensor $\BB$
leads to the equation of the type
$\DT{}\BB+
f(\BB)=(\nabla\vv)\BB+\BB(\nabla\vv)+\varepsilon\Delta\BB$
with some function $f:\R^{3\times3}\to\R^{3\times3}$, as used in
\cite{BaBuMa23CNSF,BMPS21IHCV,GaKoTr22VCHM,LMMiNe15GEUR,LMNR17GERG}, combined with the
Stokes rheology into the Jeffreys' type model.

\def\Lp{\LL\IN}
The variant of \eq{Lee-formula} in terms of the elastic left Cauchy-Green strain
$\BB\EL=\FF\EL\FF\EL^\top$ reads as 
\begin{align}
\DT{\BB\EL}=(\nabla\vv)\BB\EL+\BB\EL(\nabla\vv)-2\FF\EL({\rm sym}\LL\IN)\FF\EL^\top
\end{align}
with $\LL\IN$ as in \eq{Lee-formula}. The symmetric part is here driven 
by the inelastic flow rule replacing \eq{Max-Euler-c} as
$\Kv\,{\rm sym}\LL\IN=
\FF\EL^{-1}\varphi'(\BB\EL)\FF\EL$, possibly regularized by the gradient term
${\rm div}(\ell_2^2\Kv{\rm sym}\LL\IN)$ like in \eq{Max-Euler-c}.
Such models have been scrutinized in 
\cite{GaKoTr22VCHM,MalPru18DECM,MPSS18TVRT,MaRaTu15VMOM,PTPS22CNCF}.
For a survey see \cite{BMPS18PDEA}.

\begin{remark}[{\sl Linearized convective models.}]\label{rem-linearized}\upshape
  The full Eulerian model in terms of the nonsymmetric deformation gradient
  $\FF$ and non-quadratic stored energy $\varphi= \varphi(\FF)$ is often
  linearized by using a quadratic stored energy $\frac12(\bbC\EE){:}\EE$ in
  terms of a symmetric strain $\EE$, here $\bbC$ denotes the 4th-order
  elastic-moduli tensor. The analog of $\Dt\sigma={C}v_x$ is then
  $\ZJ\TT=\bbC\ee(\vv)$ where $\ZJ\TT:=\DT{}\TT+\TT\WW-\WW\TT$ with the spin
  tensor $\WW=\frac12\nabla\vv-\frac12(\nabla\vv)^\top$ denotes the
  corotational {\it Zaremba-Jaumann time derivative}. This determines the
  Cauchy stress $\TT$ instead of the latter equation in \eq{KV-Euler1}. 
Usage of this corotational derivative for the objective Cauchy stress rate was
justified by Biot \cite[p.494]{Biot65MID}. In the isotropic media, it is
inherited for the symmetric strain tensor $\EE$,
leading to the kinematic constraint $\ZJ\EE=\ee(\vv)$,
cf.\ \cite[Remark 2.1]{Roub23SPTC}, being the analog to $\Dt e=v_x$ in
\eq{wave-eq}. The analog to the regularization by the strain diffusion
\eq{diffusive-regularization} is then $\ZJ\EE=\ee(\vv)+\varepsilon\Delta\EE$;
for an incompressible case see \cite{BuMaMi17SDRN}.
 In the case of the Maxwell rheology, the analog of \eq{Max-small-stress} is now
 $\bbC^{-1}\ZJ\TT+\bbD^{-1}\TT=\ee(\vv)$ with $\bbD$ a 4th-order elastic-moduli
tensor, as analyzed in \cite{LioMas00GSSO} in the isotropic incompressible case.
The analog to the stress diffusion \eq{Max-stress-diffuse}
was used and analyzed in \cite{EiHoMi22LHSV} in a nonlinear incompressible
variant $\ZJ\TT+f(\TT)=\bbC\ee(\vv)+\varepsilon\Delta\SS$ combined
with the Stokes viscosity, i.e.\ actually the regularized Jeffreys rheology.
It should be noted that the combination with the ideally rigid
rheology in the volumetric case, i.e.\ the incompressible variants
with ${\rm div}\,\vv=0$, leads to the linear constraint ${\rm tr}\EE=0$
instead of the nonlinear constraint $\det\FF=1$ in the fully nonlinear models.
\end{remark}

\begin{remark}[{\sl Anomalous dispersion by conservative gradients.}]\label{rem-disp-anomalous-large+}\upshape
As in Remark~\ref{rem-normal}, 
the combination of the dissipative gradient with the gradient theory in the
conservative part like \eq{dispersion+-} here leads to the stored energy
$\varphi=\varphi(\FF)$ expanded as $\frac12\ell_2^2{C}{\mid}\nabla\FF{\mid}^2$.
Dictated by correct energetics arising from the test of the momentum
equation by $\vv$, this gradient term gives the conservative hyperstress
contribution to the Cauchy stress as
\begin{align}
\!\mathscr{S}(\FF,\nabla\FF)={\rm div}\big(\ell_2^2{C}\nabla\FF\big)\FF^\top
\ \text{ and }\ 
\mathscr{K}(\nabla\FF)=\ell_2^2{C}\Big(\nabla\FF{\otimes}\nabla\FF
-\frac12{\mid}\nabla\FF{\mid}^2\bbI\Big)\,,\!
\end{align}
respectively. Similarly as in Remark~\ref{rem-normal}, there are however
mathematical problems with this extension
particularly due to the quadratic nonlinearity of
$\mathscr{K}:\R^{3\times3}\to\R_{\rm sym}^{3\times 3}$. Let us
only vaguely remark that its analysis would
need strong convergence (of an approximation) of $\nabla\FF$,
which seems difficult unless some 3rd-grade multipolar viscosity or
some stress diffusion would be additionally involved.
\end{remark}

\section{Conclusion}

The connection between propagation of waves in linear viscoelastic media and nonlinear
models at large strains used in literature is scrutinized with a focus on 
higher-order spatial-gradient extensions of viscoelastic rheological models.
It is shown that, sometimes, higher gradients can facilitate propagation of waves
in some frequency range by lower dispersion and attenuation
than conventional models, cf.\ Figures\ \ref{KV-dispersion-comparison},
\ref{Max-dispersion-grad}, or \ref{Max-dispersion-grad+}.
Some higher-gradient extensions may compensate dispersion and result in
nondispersive models, cf.\ Figures\ \ref{KV-dispersion-anomalous},
which applies not only to a stored-energy extension but 
also to the diffusive extension of the kinematic constraints
leading to \eq{Brenner-disip+}. It is shown (at least in particular
cases in Remarks~\ref{rem-KV-hyper-enrg} and \ref{rem-KV-hyper-enrg+}
although similar phenomenon applies essentially in other models)
that, when a regularity of solutions is ensured by the regularity of
initial conditions, such higher gradients do not substantially
influence the energetics of the model if the respective length-scale
parameter $\ell$ is small.

Simultaneously, such higher gradients can facilitate rigorous analysis
at large strains either in the Lagrangian or in the Eulerian formulations,
which is usually the only motivation for such terms used in mathematically
oriented literature.
The dispersion/attenuation analysis for linear models in Section~\ref{sec3}
can vaguely offer certain additional interesting attributes for such nonlinear
large-strain models which are usually not taken into account in literature. 
For example, the dispersion analysis behind \eq{diffusive-regularization},
showing possible compensation of dispersion and easier propagation of
high-frequency waves, somehow motivates the physically unjustified diffusive
extension \eq{kinem-Euler-reg} used sometimes in literature.
 Also in other gradient models which are sometimes considered as
merely mathematically motivated and thus slightly controversial, the
dispersion/attenuation analysis in one-dimensional linear analogs may serve
as a certain argumentation for possible usage of such terms towards fitting
the models to experimentally observed phenomena. 

On the other hand, some gradient extensions do not seem rigorously
analyzed at large strains in literature, which gives challenges
for possible future research.

\bigskip

{\it Acknowledgments.}
Support from the CSF grant no.\,23-06220S and the institutional
support RVO: 61388998 (\v CR) is gratefully acknowledged.

{\small

\bigskip\bigskip\bigskip

\noindent
Mathematical Institute, Faculty of Math. \& Phys., Charles University,\\
Sokolovsk\'a 83, CZ-186~75~Praha~8,  Czech Republic,\\[.3em]
and\\[.3em]
Institute of Thermomechanics, Czech Academy of Sciences,\\
Dolej\v skova 5, CZ-18200~Praha~8, Czech Republic\\
email: ${\texttt{tomas.roubicek@mff.cuni.cz}}$

} 

\end{document}